%% file: main.tex
\documentclass[fleqn,usenatbib]{mnras}

\usepackage{newtxtext,newtxmath}

\usepackage[T1]{fontenc}

\DeclareRobustCommand{\VAN}[3]{#2}
\let\VANthebibliography\thebibliography
\def\thebibliography{\DeclareRobustCommand{\VAN}[3]{##3}\VANthebibliography}

\usepackage{graphicx}	
\usepackage{amsmath}	
\usepackage{amssymb}	
\usepackage{float}      

\title[Bayesian real-bogus classification for GOTO]{Transient-optimised real-bogus classification with Bayesian Convolutional Neural Networks --- sifting the GOTO candidate stream}

\author[Killestein et al.,]{
T. L. Killestein,$^{1}$\thanks{t.killestein@warwick.ac.uk}
J. Lyman,$^{1}$ 
D. Steeghs,$^{1}$ 
K. Ackley,$^{2}$ 
M. J. Dyer,$^{3}$ 
K. Ulaczyk,$^{1}$
\newauthor
R. Cutter,$^{1}$
Y.-L. Mong,$^{2}$ 
D. K. Galloway,$^{2}$ 
V. Dhillon,$^{3, 10}$ 
P. O'Brien,$^{4}$
G. Ramsay,$^{5}$
\newauthor
S. Poshyachinda,$^{6}$
R. Kotak,$^{7}$ 
R. P. Breton,$^{8}$ 
L. K. Nuttall,$^{9}$ 
E. Pall\'e,$^{10}$ 
D. Pollacco,$^{1}$
\newauthor
E. Thrane,$^{2}$ 
S. Aukkaravittayapun,$^{6}$
S. Awiphan,$^{6}$ 
U. Burhanudin,$^{3}$ 
P. Chote,$^{1}$
\newauthor
A. Chrimes,$^{11}$ 
E. Daw,$^{3}$ 
C. Duffy,$^{5}$ 
R. Eyles-Ferris,$^{4}$
B. Gompertz,$^{1}$ 
T. Heikkil\"a,$^{7}$
\newauthor
P. Irawati,$^{6}$ 
M. R. Kennedy,$^{8}$ 
A. Levan,$^{11,1}$
S. Littlefair,$^{3}$ 
L. Makrygianni,$^{3}$ 
\newauthor
D. Mata S\'anchez,$^{8}$
S. Mattila,$^{7}$ 
J. Maund,$^{3}$ 
J. McCormac,$^{1}$ 
D. Mkrtichian,$^{6}$ 
J. Mullaney,$^{3}$ 
\newauthor
E. Rol,$^{2}$
U. Sawangwit,$^{6}$
E. Stanway,$^{1}$ 
R. Starling,$^{4}$
P. A. Str\o m,$^{1}$
S. Tooke,$^{4}$ 
K. Wiersema,$^{1}$
\newauthor
S. C. Williams$^{7,12}$
\\ \\
$^{1}$Department of Physics, University of Warwick, Gibbet Hill Road, Coventry CV4 7AL, UK \\
$^{2}$School of Physics \& Astronomy, Monash University, Clayton VIC 3800, Australia \\
$^{3}$Department of Physics and Astronomy, University of Sheffield, Sheffield S3 7RH, UK \\
$^{4}$School of Physics \& Astronomy, University of Leicester, University Road, Leicester LE1 7RH, UK \\
$^{5}$Armagh Observatory \& Planetarium, College Hill, Armagh, BT61 9DG \\
$^{6}$National Astronomical Research Institute of Thailand, 260 Moo 4, T. Donkaew, A. Maerim, Chiangmai, 50180 Thailand \\
$^{7}$Department of Physics \& Astronomy, University of Turku, Vesilinnantie 5, Turku, FI-20014, Finland \\
$^{8}$Jodrell Bank Centre for Astrophysics, Department of Physics and Astronomy, The University of Manchester, Manchester M13 9PL, UK \\
$^{9}$University of Portsmouth, Portsmouth, PO1 3FX, UK \\
$^{10}$Instituto de Astrof'{i}sica de Canarias, E-38205 La Laguna, Tenerife, Spain \\
$^{11}$Department of Astrophysics/IMAPP, Radboud University, Nijmegen, The Netherlands \\
$^{12}$Finnish Centre for Astronomy with ESO (FINCA), Quantum, Vesilinnantie 5, University of Turku, FI-20014 Turku, Finland
}

\date{Accepted XXX. Received YYY; in original form ZZZ}

\begin{document}
\label{firstpage}
\pagerange{\pageref{firstpage}--\pageref{lastpage}}
\maketitle


\begin{abstract}
Large-scale sky surveys have played a transformative role in our understanding of astrophysical transients, only made possible by increasingly powerful machine learning-based filtering to accurately sift through the vast quantities of incoming data generated. In this paper, we present a new real-bogus classifier based on a Bayesian convolutional neural network that provides nuanced, uncertainty-aware classification of transient candidates in difference imaging, and demonstrate its application to the datastream from the GOTO wide-field optical survey. Not only are candidates assigned a well-calibrated probability of being real, but also an associated confidence that can be used to prioritise human vetting efforts and inform future model optimisation via active learning.
To fully realise the potential of this architecture, we present a fully-automated training set generation method which requires no human labelling, incorporating a novel data-driven augmentation method to significantly improve the recovery of faint and nuclear transient sources. We achieve competitive classification accuracy (FPR and FNR both below 1\%) compared against classifiers trained with fully human-labelled datasets, whilst being significantly quicker and less labour-intensive to build. This data-driven approach is uniquely scalable to the upcoming challenges and data needs of next-generation transient surveys. We make our data generation and model training codes available to the community.
\end{abstract}


\section{Introduction}
Transient astronomy seeks to identify new or variable objects in the night sky, and characterise them to learn about the underlying mechanisms that power them and govern their evolution. This variability can occur on timescales of milliseconds to years, and at luminosities ranging from stellar flares to luminous supernovae that outshine their host galaxy \citep{kulkarni12_timescales, villar17_transients}. Through observations of optical transient sources we have obtained evidence of the explosive origins of heavy elements (e.g. \citealt{gw170817_multimessenger}, \citealt{pian17_rproc}), traced the accelerating expansion of our Universe across cosmic time (e.g. \citealt{perlmutter99_sne}), and located the faint counterparts of some of the most distant and energetic astrophysical events known: gamma-ray bursts (e.g. \citealt{tanvir09_grb}). Requiring multiple observations of the same sky area to detect variability, transient surveys naturally generate vast quantities of data that require processing, filtering, and classification -- this has driven the development of increasingly powerful techniques bolstered by machine learning to meet the demands of these projects.

Many of the earliest prototypical transient surveys began as galaxy-targeted searches, performed with small field-of-view instruments. In the early stages of these surveys candidate identification was performed manually, with humans `blinking' images to look for varying sources. This process is time-consuming and error-prone, and represented a bottleneck in the survey dataflow which heavily limited the sky coverage of these surveys. The first `modern' transient surveys (e.g. LOSS; \citealt{lick_snsearch}) used early forms of difference imaging to detect candidates in the survey data, automating the candidate detection process and enabling both faster response times and greater sky coverage. LOSS proved extremely successful, discovering over 700 supernovae in the first decade of operation, providing a homogeneous sample that has proven useful in constraining supernova rates for the local Universe \citep{leaman11_licksnrates, li11_loss2}. 

Difference imaging has since emerged as the dominant method for the identification of new sources in optical survey data. With this method, an input image has a historic reference image subtracted to remove static, unvarying sources. Transient sources in this difference image appear as residual flux, which can be detected and measured photometrically using standard techniques.  Various algorithms have been proposed for optical image subtraction, either attempting to match the point spread function (PSF) and spatially-varying background between an input and reference image \citep{alard98_diffimg, becker15_hotpants}, or accounting for the mismatch statistically \citep{zackay16_zogy} to enable clean subtraction. Difference imaging also provides an effective way to robustly discover and measure variable sources in crowded fields \citep{wozniak00_diffimg}.

Driven by both improvements in technology (large-format CCDs, wide-field telescopes) and difference imaging algorithms, large-scale synoptic sky surveys came to the fore. In this mode, significant areas of sky can be covered each night to a useful depth and candidate transient sources automatically flagged. This has driven an exponential growth in discoveries of transients, with over 18,000 discovered in 2019 alone\footnote{\url{https://wis-tns.weizmann.ac.il/}}. Wide-field surveys such as the Zwicky Transient Facility (ZTF; \citealt{ztfprojectpaper19}), PanSTARRS1 (PS1; \citealt{ps1_projectpaper16}), the Asteroid Terrestrial-impact Last Alert System (ATLAS; \citealt{atlas_projectpaper18}), and the All Sky Automated Survey for SuperNovae (ASAS-SN;  \citealt{shappee14}) have proven to be transformative, collectively discovering hundreds of new transients per night.

With the ability to repeatedly and rapidly tile large areas of sky in order to search for new and varying sources, the follow-up of optical counterparts to poorly localised external triggers became possible, in the process ushering in the age of multi-messenger astronomy. An early example was detection of optical counterparts to {\it Fermi} gamma-ray bursts by the Palomar Transient Factory (PTF; \citealt{ptf_projectpaper09}). Typical localisation regions from the {\it Fermi} GBM instrument \citep{meegan09_fermigbm} were of order 100 square degrees at this time, representing a significant challenge to successfully locate comparatively faint ($r \sim 17-19$) GRB afterglows. Of the 35 high-energy triggers responded to, 8 were located in the optical \citep{singer15_fermi}, demonstrating the emerging effectiveness of synoptic sky surveys for this work.

Another recent highlight has been the detection of an optical counterpart to a TeV-scale astrophysical neutrino detected by the IceCUBE facility \citep{icecube17_projectpaper}. Recent and historical wide-field optical observations of the localisation area combined with high-energy constraints from {\it Fermi} enabled the identification of a flaring blazar, believed to be responsible for the alert (IceCube-170922A; \citealt{icecube18_blazar})  . This rapidly increasing survey capability has culminated recently in the landmark discovery of a multi-messenger counterpart to the gravitational wave (GW) event GW170817 \citep{gw170817_gwonly, gw170817_multimessenger}.  

\subsection{Real-bogus classification} \label{sec:realbogus_overview}
For many years, the rate of difference image detections generated per night by sky surveys has significantly exceeded the capacity of teams of humans to manually vet and investigate each one. This has motivated the development of algorithmic filtering on new sources, to reject the most obvious false positives and reduce the incoming datastream to something tractable by human vetting. With the growing scale and depth of modern sky surveys, simple static cuts on source parameters cannot keep pace with the rate of candidates, with high false positive rates leading to substantial contamination by artifacts. This situation has motivated the development of machine learning (ML) and deep learning (DL) classifiers, which can extract subtle relationships/connections between the input data/features and perform more effective filtering of candidates. The dominant paradigm for this task has so far been the real-bogus formalism \citep[e.g.][]{bloom12}, which formulates this filtering as a binary classification problem. Genuine astrophysical transients are designated `real' (score 1), whereas detector artefacts, subtraction residuals and other distractors are labelled as `bogus' (score 0). A machine learning classifier can then be trained using these labels with an appropriate set of inputs to make predictions about the nature of a previously-unseen (by the classifier) source within an image.

This real-bogus classification is only one step in a transient detection pipeline. Having established the candidates appearing as astrophysically real sources, further filtering is required to determine if they are scientifically interesting, or distractors -- the definition of ``interesting'' is naturally governed by the science goals of the survey. This process draws in contextual information from existing catalogues, historical evolution, and more fine-grained classification routines.  The last step before triggering follow-up and further study (at least currently) is human inspection of the remaining candidates. No single filtering step is 100\% efficient in removing false positives/low significance detections, thus human vetting is required to identify promising candidates and screen out any bogus detections that have made it this far. Real-bogus classification is the most crucial step, reducing the volume of candidates that later steps must process and the amount of bogus candidates that humans must eventually sift through to find interesting objects -- a balance between sensitivity (to avoid missing detections irretrievably) and specificity (avoiding floods of low-quality candidates) must be reached.

Real-bogus classification is a well-studied problem, beginning with early transient surveys \citep{romano06, bailey07}, and evolving both in complexity and performance with the increasing demands placed on it by larger and deeper sky surveys such as PTF \citep{brink13}, PanSTARRS1 \citep{ps1_projectpaper16}, and the Dark Energy Survey \citep{goldstein15}. Early classifiers were generally built on decision tree-based predictors such as random forests \citep{breiman01}, using a feature vector as input. Feature vectors comprise extracted information about a given candidate, and often include broad image-level statistics/descriptions designed to maximally separate real and bogus detections in the feature space. Examples include the source full-width half maximum computed from the 2D profile, noise levels, and negative pixel counts. More elaborate features can be composed via linear combinations of these quantities, which may exploit correlations and symmetries. Another method of deriving features is to compute compressed numerical representations of the source via Zernicke/shapelet decomposition \citep{ackley19_zernicke}.

However, feature selection can represent a bottleneck to increasing performance. Features are typically selected by humans to encode the salient details of a given detection, attempting to find a compromise between classification accuracy and speed of evaluation. This introduces the possibility of missing salient features entirely, or choosing a sub-optimal combination of them. 

Directly using pixel intensities as a feature representation avoids choosing features entirely, instead training on flattened and normalised input images \citep{wright15, mong20}, these have demonstrated improved accuracy over fixed-feature classifiers. However, this approach quickly (quadratically) becomes inefficient for large inputs. Using a smaller input size means information on the surrounding area of each detection is unavailable, limiting the visible context and affecting classification accuracy as a result. 

Recently, convolutional neural networks (CNNs, \citealt{lecun95}) have led to a paradigm shift in the field of computer vision and machine learning, which has been transformative in the way we process, analyse, and classify image data across all disciplines. CNNs use learnable convolutional filters known as kernels to replace feature selection. These filters are cross-correlated with the input images to generate `feature maps', effectively compact feature representations. Through the training process, the filter parameters are optimised to extract the most salient details of the inputs, which can then be fed into fully-connected layers to perform classification or regression. In this way, the model can select its own feature representations, avoiding the bottleneck of human selection. Multiple layers can be combined to achieve greater representational power, known as deep learning \citep{lecun15}. Recent work using CNNs has demonstrated state-of-the-art performance at real-bogus classification \citep{gieseke17, cabreravives17_deephits, duev19, turpin20}.  CNNs are also efficiently parallelisable making them suitable for high-volume data processing tasks. Whilst providing substantial accuracy improvements over previous techniques, deep learning is particularly reliant upon large and high quality training sets to minimise overfitting, arising from the high number of model parameters. Although augmentation and regularisation techniques can minimise this risk, they are no substitute for a larger dataset. The performance of any classifier is ultimately limited by the error rate on the training labels, so it is important to also ensure the dataset is accurately labelled. Making a large, pure, and diverse training set can be among the most challenging parts of developing a machine learning algorithm, and significant effort has been focused on this area in recent years.

Traditionally the `gold-standard' for machine learning datasets across computer science and astronomy has been human-labelled data, as this represents the ground truth for any supervised learning task. Use of citizen science has proven to be particularly effective, leveraging large numbers of participants and ensembling their individual classifications to provide higher accuracy training sets for machine learning through collaborative schemes such as Zooniverse \citep{lintott08_galaxyzoo, mahabal19_ztf}. However, even in large teams, human labelling of large-scale datasets is time-consuming and inefficient requiring hundreds--thousands of hours spent collectively to build a dataset of a suitable size and purity. Specifically for real-bogus classification, there are also issues with completeness and accuracy for human labelling of very faint transients close to the detection limit. These faint transients are where a classifier has potential to be the most helpful, so if the training set is fundamentally biased in this regime, any classifier predictions will be similarly limited. To go beyond human-level performance, we cannot solely rely on human labelling, additional information is required. One specific aspect of astronomical datasets that can be leveraged to address both issues discussed above is the availability of a diverse range of contextual data about a given source. Sizeable catalogues of known variable stars, galaxies, high energy sources, asteroids, and many other astronomical objects are freely available and can be queried directly to identify and provide a more complete picture of the nature of a given source. 

Significant effort is being invested in data processing techniques for transient astronomy in anticipation of the Vera C. Rubin Observatory \citep{ivezic19_lsst}, due to begin survey operations in 2022. Via the Legacy Survey of Space and Time (LSST), the entire southern sky will be surveyed down to a depth of $r' \sim 24.5$ in 5 colours at high cadence, providing an unprecedented discovery engine for transients to depths previously unprobed at this scale. The dataflow from this project is expected to be a factor 10 greater than current transient surveys, and promises to be transformative in the fields of supernova cosmology, detection of potentially hazardous near-Earth asteroids, and mapping the Milky Way in unprecedented detail. The main high-cadence deep sky survey promises to provide a significant increase in the number of genuine transients we detect, but also a significant increase in the number of bogus detections assuming there are not similarly large improvements in the capability of machine learning-based filtering techniques. Development of higher-performance classifiers is crucial to fully exploit this stream, but also more granular classification involving contextual data (as recently demonstrated by \citealt{alerce20}) to ensure that novel and scientifically important candidates are identified promptly enough to be propagated to teams of humans and followed up. 

A related goal of increasing importance in the big data age of the Rubin Observatory and similar projects is that of quantifying uncertainty -- being able to identify detections that the classifier is confident are real, and providing a classifier a way to indicate uncertainty on more tenuous examples. This objective goes beyond the simple value of the real-bogus score, and can then be used to find the optimal edge cases to feed to human labellers, allowing new data to be continually integrated to improve performance and keep the classifier's knowledge current and applicable to a continuously evolving set of instrumental parameters. Current generation transient surveys provide a crucial proving ground for development of these new techniques.

\subsection{The Gravitational-Wave Optical Transient Observer (GOTO)}
The Gravitational-Wave Optical Transient Observer \citep{goto_prototype_paper} is a wide-field optical array, designed specifically to rapidly survey large areas of sky in search of the weak kilonovae and afterglows associated with gravitational wave counterparts. The work we present in this paper was conducted during the GOTO prototype stage, using data taken with a single `node' of telescopes situated at the Roque de los Muchachos observatory on La Palma. Each node comprises 8 co-mounted fast astrograph OTAs (optical tube assemblies) combining to give a $\sim 40$ square degree field of view in a single pointing. GOTO performs surveys using a custom wide $L$ band filter (approximately equivalent to $g' + r'$) down to $L \approx 20$, providing an effective combination of fast and deep survey capability uniquely suited to tackling the challenging large error boxes associated with gravitational wave detections. As demonstrated in \cite{gompertz20_gotobbh}, the prototype GOTO installation is capable  of conducting sensitive searches for the optical counterparts of nearby binary neutron star mergers, even with weak localisations of $\sim$1000 square degrees. When not responding to GW events, GOTO performs an all-sky survey utilising difference imaging to search for other interesting transient sources. Although the GOTO prototype datastream will be the primary data source used to investigate the performance of the machine learning techniques developed in this paper, the methods are inherently scalable and will also be deployed for the future GOTO datastream from 4 nodes spread over two sites. For now, we concentrate on a calendar year of prototype operations (spanning 01-01-2019 -- 01-01-2020) -- which represents a significant dataset, comprising 44,789 difference images in total.

Raw images are reduced with the GOTO pipeline \citep{goto_prototype_paper}. Here we provide a very brief overview of the process for context, and delegate more in-depth discussion to the specific upcoming pipeline papers. The typical survey strategy for GOTO is three exposures per pointing, which undergo standard bias, dark and flat correction, and then are median-combined to reject artifacts and improve depth. Throughout this paper we refer to this median-combined stack of subframes as a `science image'.
Each combined image is matched to a reference template, which passes basic quality checks, and aligned using the {\sc spalipy}\footnote{\url{https://github.com/Lyalpha/spalipy}} code. Image subtraction is performed on the aligned science and reference images with the {\sc hotpants} algorithm \citep{becker15_hotpants} to generate a difference image. To locate residual sources in the difference image, source extraction is performed using {\sc SExtractor} \citep{bertin96_sextractor}. Detections in the difference image are referred to as `candidates' through the remainder of this paper. For each candidate, a set of small stamps are cut out from the main science, template and difference images and this forms the input to the GOTO real-bogus classifier. This process and proposed improvements are discussed in more detail in Section \ref{sec:stamps}. 
From here, candidates that pass a cut on real-bogus score (using a preliminary classifier) are ingested into the GOTO Marshall -- a central website for GOTO collaborators to vet, search and follow-up candidates (Lyman et al., in prep.).

In line with the principal science goals of the GOTO project, the real-bogus classifier discussed in this work is constructed specifically to maximise the recovery rate of extragalactic transients and other explosive events such as cataclysmic variable outbursts. Small-scale stellar variability can be easily detected via difference imaging, but is better studied through the aggregated source light curves. An operational requirement for the current version of this classifier is the ability to perform consistently across multiple different hardware configurations. During classifier development, the GOTO prototype used two different types of optical tube design, each with varying optical characteristics that led to different point spread functions, distortion patterns, and background levels/patterns. Due to limited data availability, training a classifier for each individual OTA (or group of OTAs of the same type) was not viable. This requirement adds an additional operational challenge over survey programs such as the Zwicky Transient Facility \citep[ZTF,][]{ztfprojectpaper19} and PanSTARRS1 \citep[PS1,][]{ps1_projectpaper16}, which use a static, single-telescope design. If acceptable results can be achieved with this heterogeneous hardware configuration, then further performance gains can be expected when the design GOTO hardware configuration is deployed. This will use telescopes of consistent design and improved optical quality meaning less model capacity needs to be directed towards making the classification performance stable and across a diverse ensemble of optical distortions.

In this paper, we propose an automated training set generation procedure that enables large, minimally contaminated, and diverse datasets to be produced in less time than human labelling and at larger scales. This procedure also introduces a data-driven augmentation scheme to generate synthetic training data that can be used to significantly improve the performance of any classifier on extragalactic transients of all types, but with particular effectiveness for nuclear transients. Using this improved training data, we apply Bayesian convolutional neural networks (BCNNs) to astronomical real-bogus classification for the first time, providing uncertainty-aware predictions that measure classifier confidence, in addition to the typical real-bogus score. This opens up promising future directions for more complex classification tasks, as well as optimally utilising the predictions of human labellers. We emphasise that although this classifier is discussed in the context of GOTO and our associated science needs, the techniques discussed are fully general and could be applied to general real-bogus classification at other projects easily. Our code, {\sc gotorb}, is made freely available online \footnote{\url{https://github.com/GOTO-OBS/gotorb}} with this in mind.

\section{Training set generation and augmentation}
The `real' content of our training set is composed of minor planets, similar to \citet{smith20_atlassciserv}. Assuming the sky motion is large (but not so large that the source is trailed) these objects are typically detected in the science image but not the template image, which provides a clean subtraction residual resembling an explosive transient. Due to the large pixels of the GOTO detectors and short exposure times of each sub-image, very few asteroids move sufficiently quickly to trail. We estimate that sky motions of 1 arcsec per minute or greater will lead to trailing. 

There are significant numbers of asteroids detectable down to $L \sim 20.5$ with GOTO, and the sky motion ensures that a diverse range of image configurations are sampled. With the large $\sim 40$ square degree field of view provided by GOTO, an whole-sky average of 4.6 asteroids per pointing are obtained, with this number significantly increasing towards the ecliptic plane. Using ephemerides provided by the \texttt{astorb} database \citep{astorb19}, based on observations reported to the Minor Planet Center\footnote{\url{https://www.minorplanetcenter.net/}}, difference image detections can be robustly cross-matched to minor planets in the field. This provides a significant pool of high-confidence, unique, and diverse difference image detections from which to build a clean training set.

\begin{figure}
    \centering
    \includegraphics[width=\columnwidth]{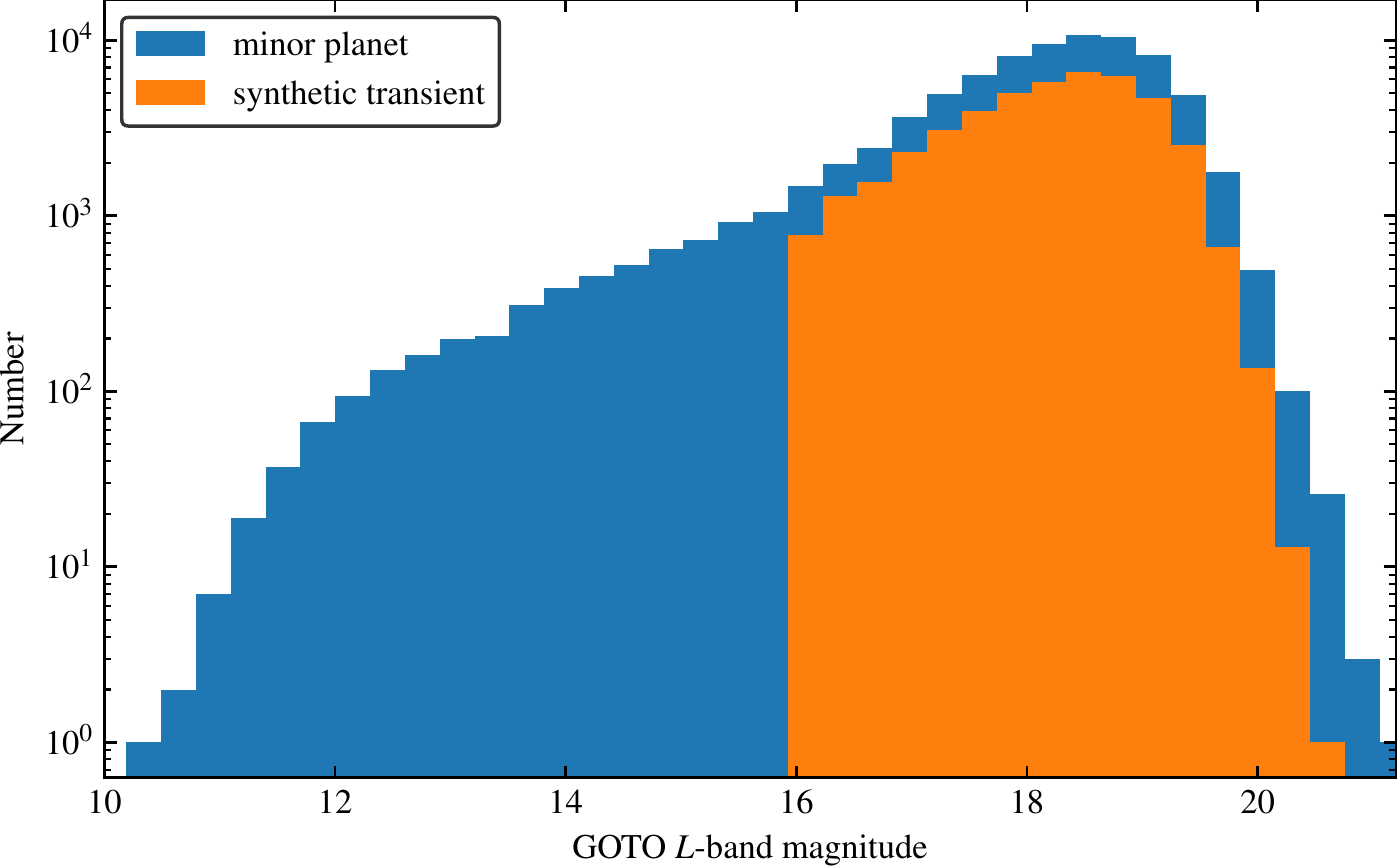}
    \caption{Magnitude distribution of the minor planets (MP) used to build our training set. Bright-end number densities are dominated by the true magnitude distribution of the minor planets, where the faint-end density is constrained by the GOTO limiting magnitude. The magnitude distribution of synthetic transients (SYN) is a sub-sample of the minor planet magnitude distribution, except with a cut at $L\sim$16, to avoid unrealistically bright objects.}
    \label{fig:magdist}
\end{figure}

We use the online SkyBoT cone search \citep{berthier06, berthier16} to retrieve the positions and magnitudes of all minor planets within the field of view of each GOTO image, then cross-match this table with all valid difference image detections using a 1 arcsec threshold value to identify the asteroids present in the image. The ephemerides provided are of sufficient quality that this is adequate to match even faint ($L \sim$ 20) asteroids. To avoid spurious cross-matches, only asteroids brighter than the 5-sigma limiting magnitude of the image are considered. An alternative offline cone search is made accessible via the {\sc pympc} package\footnote{\url{https://pypi.org/project/pympc/}} Python package, which the code can fall back on if SkyBoT is unavailable. Using minor planets, the training set can reliably be extended to fainter magnitudes, where the performance of human vetters begins to significantly decrease. Figure \ref{fig:magdist} illustrates the magnitude distribution of minor planets used to construct the training set. 

To create the bogus content of our training set, we randomly sample detections in the difference image following \cite{brink13}. Bogus detections overwhelmingly ($\gtrsim 99\%$) outnumber real detections in each difference image, so it is justified to sample in this way.  One significant source of contamination taking this approach is variable stars, therefore we remove all known variable stars from the random bogus component by cross-matching against the ATLAS Variable Star Catalogue \citep{heinze18varcat} with a 5 arcsec radius. These variable star detections can constitute 2--4\% of the entire bogus dataset. Of the detections removed by this step, a small fraction of these will be high-amplitude variable stars which have a strong subtraction residual in a given night's data, and thus represent real sources lost. Automating the correct labelling of these sources using light curve information is feasible, but would add significant complexity and more potential failure modes, so we instead opt to remove the variable stars entirely and simply add more verifiably `real' detections in their place in the form of more minor planets. Inevitably, some small fraction of uncatalogued variable stars will be missed with this procedure, and we develop tools to identify them retrospectively after model training in Section \ref{sec:bayesian_tricks}.

To improve the classifier's resistance to specific challenging subtypes of data poorly represented in our algorithmically generated training set, we inject human-labelled detections into the dataset. More specifically, candidates from the GOTO Marshall (discussed in full in Lyman et al., in prep.) are included, which were misidentified by the classifier in the pipeline at the time as real and later labelled as bogus by human vetters. The previous classifier was a rapidly-deployed prototype CNN similar in design to that presented here, trained on a smaller dataset of minor planets and random bogus detections. These detections are included to allow the classifier to screen out artifacts missed by the prototype image processing pipeline, including satellite trails and highly wind-shaken PSFs. This artifically increases the diversity of the bogus component of the training set, as these edge-case detections would rarely be selected by naive random sampling and so be poorly represented within the model. Although these detections represent a small fraction of the overall training set ($\sim$ 5\%), they provide a marked improvement in performance in the real-world deployment of the classifier, including marginal gains on more typical detections.

\subsection{Data extraction and format} \label{sec:stamps}
\begin{figure}
    \centering
    \includegraphics[width=\columnwidth]{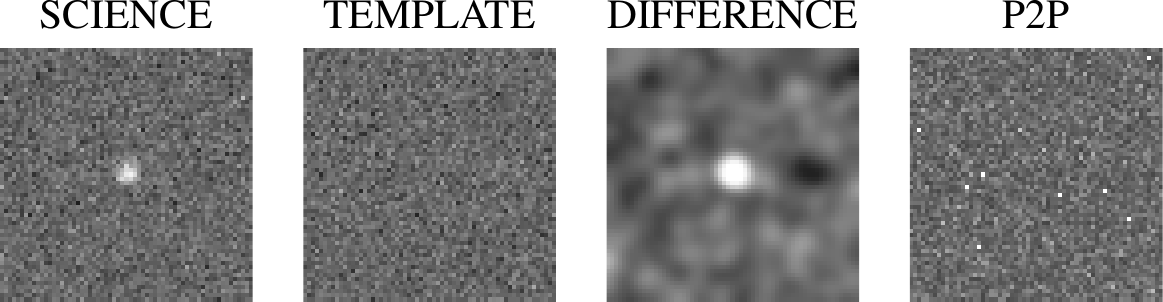}
    \caption{Example data format for a set of idealised synthetic images of a single Gaussian source newly appearing in the science image.  We apply a naive convolution of science image with template PSF and vice versa in producing the difference image for visualisation purposes.
    From left to right: science median, template median, difference image, pixel-wise peak-to-peak variation across contributing images to science median. Cutouts are 55x55 pixels square, corresponding to a side length of 1.1 arcminutes.}
    \label{fig:trainingsetcomposition}
\end{figure}

For each detection identified for inclusion in our training/validation/test sets, a series of stamps are cut out from the larger GOTO image centred on the difference image residual. In common with previous CNN-based classifiers, we use small cutouts of the median-stacked science and template images, as well as the resultant difference image after image subtraction. The size of these stamps is an important model hyperparameter, which we explore in more detail in Section \ref{sec:hyperparams}. A example of the model inputs for a synthetic source are illustrated in Figure \ref{fig:trainingsetcomposition}.

An important addition to our network's inputs compared to previous work is a peak-to-peak (\texttt{p2p}) layer. This is included to characterise variability across the individual images that make up a median stacked science image, and is calculated as the peak-to-peak  (maximum value - minimum value) variation of each pixel computed across all individual images that composed the median stack. To ensure consistent alignment across all individual stamps and remove any jitter, we cut out the region based on the RA/Dec coordinates of the source detection in the median stack. This additional provides an effective discriminator for spurious transient events such as cosmic ray hits and satellite trails. If sufficiently bright, these are not removed by the simple median stacking in the current pipeline due to the small number of sub-frames used. This is particularly problematic for cosmic ray hits which are convolved with a Gaussian kernel for image subtraction, and appear PSF-like in the difference image. This can create convincing artifacts which are difficult to identify without access to the individual image level information. In testing, this reduced the false positive rate on the test set by  $\sim$ 0.2\%. Although this is not a sizeable improvement when evaluated on the full dataset, cosmic ray hits constitute a very small percentage of overall detections. Testing instead on a human-labelled set of bogus detections which were initially scored as real by the existing deployed classifier (without a {\tt p2p} layer), there is a 2--3\% decrease in false positive rate. 

For all of the above steps, stamps extending beyond the edge of the detector have missing areas filled in with a constant intensity level of $10^{-6}$, to distinguish them quantitatively from masked (i.e. saturated) pixels which are assigned a value of zero in the difference image by the pipeline. The specific intensity level chosen for this offsetting is not important, and we choose our value to be well above machine precision (significant enough to influence the gradients) but well below the typical background level.
To ensure that the classifier remains numerically stable in later training steps, each stack of stamps undergoes layer-wise L2 normalisation to reduce the input's magnitude. Each stamp has the mean subtracted and is then divided through by the L2 ($\sqrt{\vec{x}
\cdot \vec{x}}$) norm.

\subsection{Synthetic transients} \label{sec:syntransients}
Although asteroids provide a convenient source of PSF-like residuals to train on, it is important to note that they cannot fully replicate genuine transients. Asteroids are markedly simpler to learn and discriminate for a classifier since they lack the complex background of a host galaxy. The main goal of this classifier is to detect extragalactic transients, so adapting the training set to maximise performance on these objects is important.
An ideal approach would be to add a large number of genuine transients into the training set. However, GOTO has not been on-sky long enough to collect a suitably large set of these detections, and we only build the training set from the previous year of data. Even assuming every supernova over the past year is robustly detected in our data this will still yield a number of transients that is significantly less than the target size of our training set. This would create a severely imbalanced dataset, which could in principle be used but with reduced classification performance.
Using spectroscopically confirmed transients may also inject an element of observational bias into our training set, as events that have favourable properties for spectroscopy (in nearby galaxies, offset from their host, bright) are preferentially selected \citep{bloom12} to be followed up. Instead we reserve a set of real, spectroscopically confirmed transients GOTO has detected ($\sim 900$ as of August 2020) for benchmarking purposes, as they represent a valuable insight into real-world performance and can be used to directly evaluate the effectiveness of any transient augmentation scheme we employ, as in Section \ref{sec:spectrans}.

PSF injection has been used heavily in prior work to generate synthetic detections for testing recovery rates and simulating the feasibility of observations. This process can be computationally intensive, involving construction of an effective PSF (ePSF) from combining multiple isolated sources or fitting an approximating function (e.g. a Gaussian) to sources in the image. The ePSF model can then be scaled and injected into to the image to simulate a new source. By injecting sources in close proximity to galaxies in individual images then propagating this through the data reduction pipeline, synthetic transients could be generated in a realistic way. However, the fast optical design of GOTO makes this a complex task, as the PSF varies as a function of source position on the detector. Sources in the corners of an image display mild coma, which, combined with wind-shake and other optical distortion, can lead to unusual PSFs that are not accurately reproduced by the mean PSF. In principle this could be accounted for by computing PSFs for sub-regions of a given image or assuming some spatially-varying kernel to fit for, but this would add sizeable overheads to the injection process and will always be an approximation.

Recent new techniques such as generative adversarial networks (GANs, \citealt{goodfellow14}) have shown promise in generating novel training examples that can be used to address class imbalances/scarcity in training sets \citep{mariani18}, and have recently started to be applied to astrophysical problems \citep{yip19}. However these networks are computationally expensive, complex to train and understand the outputs of, and don't fully remove the need for large datasets. A robust human-interpretable method for generating synthetic examples is a better approach for the noisy, diverse datasets used in real-bogus classification.

We propose a novel technique for synthesising realistic transients that can be used to significantly improve transient-specific performance when compared to a pure minor planet training set, without requiring PSF injection or other CPU-intensive approaches. 
For each minor planet detected in an image, the GLADE galaxy catalogue \citep{dalya18glade} is queried for nearby galaxies within a set angular distance of 10 arcminutes, chosen such that the PSF of sources within this region are consistent. Pre-built indices are used via {\sc catsHTM} \citep{soumagnac18} to accelerate querying GLADE. The algorithm chooses the galaxy with the brightest galaxy (minimum $B$ band magnitude) within range, then generates a cutout stamp with with a randomly chosen $x, y$ offset relative to the galaxy centre. For the implementation within this work, the $x, y$ pixel offsets are drawn from a uniform distribution $U(-7. 7)$ chosen to fully cover the range of offsets for nearby galaxies. Sources that are completely detached from any host galaxy are better represented by the minor planet component of the training set. This ensures that a diverse range of transient configurations (nuclear, offset, orphaned) are sampled. The minor planet and galaxy stamp are then directly summed to produce the synthetic transient. For the purposes of real-bogus classification, accurately matching the measured transient host-offset distribution is not crucial. The host offset distribution contains implicit and difficult to quantify biases resulting from the specific selection functions of the transient surveys that populate it -- it does not reflect accurately the underlying distribution of astrophysical transients. By choosing from a uniform distribution, we instead aim to attain consistent performance across a wide range of host offsets that overlap with the range inferred from the transient host offset distribution.

\begin{figure}
    \centering
    \includegraphics[width=\columnwidth]{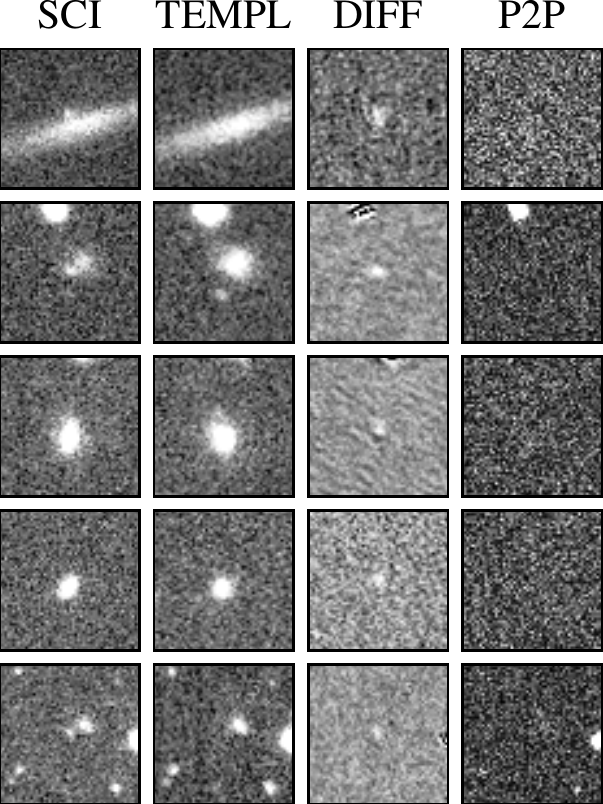}
    \caption{Randomly selected sample of synthetic transients generated with our algorithm, displayed in the same format as in Figure \ref{fig:trainingsetcomposition}. Significant variations in the PSF are visible due to sampling directly from the image, improving classifier resilience.}
    \label{fig:syntransient_examples}
\end{figure}

The original individual images for each component are retrieved to correctly compute the peak-to-peak variation of the combined stamp. Model inputs are pre-processed and undergo L2 normalisation (as discussed in Section \ref{sec:stamps}) prior to training and inference, so additional background flux introduced by this method does not affect the model inputs. The noise characteristic of this combined stamp is not straightforward to compute due to the highly correlated noise present in the difference image and varying intensity levels, and could be higher or lower depending on the specific stamps -- with the straightforward Gaussian case providing a $\sqrt{2}$ reduction in noise. This is likely not problematic for the classifier, providing a form of regularisation that could improve generalisation accuracy. We also assume that the spatial gradients in background across both stamps are $\sim$ constant, as the stamp scale is far smaller than the overall frame scale -- naturally this breaks down in the presence of nebulosity/galaxy light but this represents a overwhelmingly small fraction of the sky. We also reject all minor planets with $L < 16$, as these are significantly brighter than the selected host galaxy so are better represented by the pure minor planet candidates. This also cuts down significantly on saturated detections of dubious quality. This choice has no detrimental effect on bright-end performance, as discussed in Section \ref{sec:performance}. A random sample of synthetic transients generated with this approach is shown in Figure \ref{fig:syntransient_examples}. Our method bears some similarity in retrospect to the approach of \citep{cabreravives17_deephits}, who added stamps from the science image into difference images to simulate detections in `random' locations. Our approach uses confirmed difference image detections of MPs and puts them in more purposeful locations, whilst preserving the noise characteristics of the difference stamp.

This approach has strong advantages over simply injecting transients into galaxies. By selecting only galaxies close to each minor planet, the PSF is preserved and is consistent, regardless of how distorted it may be. Injection-based methods require estimation/assumption of the image PSF, which is typically a parameterised function determined by fitting isolated sources. Given the variation in PSF across images and across individual unit telescopes, this would be a computationally intensive task, and would likely lead to poorer results compared to using minor planets.
However, using only these synthetic transients introduces unintended behaviour in the trained model that significantly degrades classification performance if not remedied. Since every synthetic transient in the training set is associated with a host galaxy by design, the model will over time learn to associate all detections with galaxies as being real as there is no loss penalty for doing so. To resolve this, we also inject galaxy residuals as bogus detections, randomly sampling from the remaining GLADE catalog matches at a 1:1 transient:galaxy residual ratio. This way, the model learns that the salient features of these detections are not the galaxy, but the PSF-like detection embedded in them.

\subsection{Training set construction} \label{sec:trainingset}
Using the techniques developed in the sections above, we build our training set with GOTO prototype data from 01-01-2019 to 01-01-2020. This ensures that our performance generalises well across a range of possible conditions -- with PSF shape and limiting magnitude being the most important properties that benefit from this randomisation. A breakdown of training set proportions and properties is given in Table \ref{tab:datadist}.

Our code is fully parallelised at image level, meaning that a full training set of $\sim$400,000 items can be constructed in under 24h on a 32-core machine. Training sets can also be easily accumulated on multiple machines and then combined thanks to the use of the HDF5 file format. The main bottlenecks of training set generation are IO-related -- loading in image data to prepare the stamps, and querying the GLADE catalogue and SkyBoT cone search.

\begin{table}
    \centering
    \begin{tabular}{l c c c}
    Metalabel & Train & Test \\
    \hline
    \hline
    Minor planet & 72992 & 8133 & \\
    Synthetic transient & 40192 & 4521 & \\
    Random bogus & 177556 & 19645 & \\
    Galaxy residual & 28040 & 3190 & \\
    Marshall bogus & 24577 & 2662 & \\
    \hline
    Total & 343357 & 38151 & {\bf 381508}\\
    \hline
    \end{tabular}
    \caption{Breakdown of the composition of our dataset, partitioned according to training and test sets. The validation dataset is not shown, but is composed of 10\% of the training dataset, chosen randomly at training time.}
    \label{tab:datadist}
\end{table}

\section{Classifier architecture}
\begin{figure*}
    \centering
    \def\svgwidth{\linewidth}
    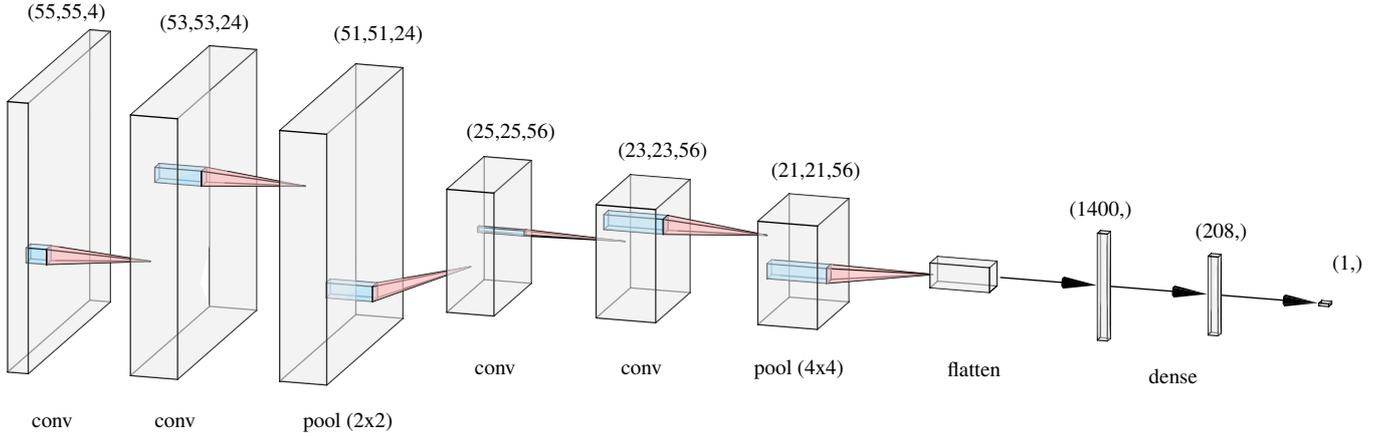
    \caption{Block schematic of the optimal neural network architecture found by hyperparameter optimisation in Section \ref{sec:hyperparams}. Each block here represents a 3D image tensor, either as input to the network, or the product of a convolution operation generating an `activation map'. Classification is performed using the scalar output of the neural network. Directly above each 3D tensor block the dimensions in pixels are shown, along with the operation that generates the next block below it represented by the coloured arrow.
    Not illustrated for clarity here are the dropout masks applied between each layer and the activation layers.
    Base figure produced with \textsc{nnsvg} \citep{lenail19}.
    }
    \label{fig:networkarch}
\end{figure*}

As a starting point, we follow the {\sc braai} classifier of \citet{duev19} in using a downsized version of the VGG16 CNN architecture of \citet{simonyan14}. This network architecture has proven to be very capable across a variety of machine learning tasks, and is a relatively simple architecture to implement and tweak. 
This architecture uses conv-conv-pool blocks as the primary component -- two convolutions are applied in sequence to extract both simple and compound features, then the resultant feature map is reduced in size by a factor 2 by `pooling', taking the maximum value of each 2x2 group of pixels. This architecture also uses small kernels (3x3) for performance. These structures are illustrated in Figure \ref{fig:networkarch}. We use the configuration as presented in \citet{duev19} for development, but later conduct a large-scale hyperparameter search to fine-tune the performance to our specific dataset (Section \ref{sec:hyperparams}). The primary inputs to the classifier are small cut-outs of the science, template, difference, and \texttt{p2p} images as discussed in Section \ref{sec:stamps} which we refer to as stamps.

The sample weights for real and bogus examples are adjusted to account for the class imbalance in our dataset, set to the reciprocal of the number of examples with each label. Class weights are not adjusted on a per-batch basis, as our training set is only mildly imbalanced.
For regularisation, we apply a penalty to the loss based on the L2 norm of each weight matrix. This penalises exploding gradients and promotes stability in the training phase. L1 regularisation was trialled but did not produce significantly better results. We also use spatial dropout \citep{tompson15} between all convolutions which provides some regularisation, but primarily is used for the purposes of uncertainty estimation (see Section \ref{sec:bayesian_tricks}) -- a small dropout probability of $\sim$ 0.01 is found to be optimal from work in Section \ref{sec:hyperparams}. Due to the significant training set size and our use of augmentation, very little regularisation is needed for a model of this (comparatively) low complexity. 

To further increase the effective size of our training set we randomly augment training examples with horizontal and vertical flips, which provide a factor 4 increase in effective training set size over unaugmented stamps. 
We also trialled the usage of 90 degree rotations following \citep{dielemann15_rotconv}, which do not require interpolations and thus do not introduce spurious artifacts that could add additional learning complexity. In constrast to other works \citep{cabreravives17_deephits, reyes18_lrp}, we find consistent performance (over multiple training runs) with simple reflections -- potentially having already reached the saturation region of the learning curve.

Our model is implemented with the {\sc Keras} framework \citep{chollet2015}, running with an optimised build of the {\sc TensorFlow} backend \citep{tensorflow2015}. For parameter optimisation we use the {\sc Adam} optimiser of \cite{kingma14}, which provides reliable convergence, and use the binary cross-entropy as the loss function. To avoid overfitting, we utilise an early stopping criterion conditioned on the validation dataset loss --- if there has been no decrease in validation loss within 10 epochs, the model training is terminated. We perform model training and inferencing on CPU only, to mirror the deployment architecture used in the main GOTO pipeline. Using a single 32-core compute node, training the finalised model to early-stopping at $\sim$170 epochs takes around 10 hours. Inferencing is significantly quicker, with an average throughput of 7,500 candidates per second with no model ensembling performed. Our model training code is freely available via the \texttt{gotorb} Python package \footnote{\url{https://github.com/GOTO-OBS/gotorb}}, which includes the full range of tunable parameters and model optimisations we implement.

\subsection{Tuning of hyperparameters/training set composition} \label{sec:hyperparams}
To achieve the maximum performance possible with a given neural network, we conduct a search over the model hyperparameters to assess which combinations lead to the best classification accuracy and model throughput.  Initially the ROC-AUC score \citep{fawcett06_roc} was used as the metric to optimise as in many cases this is a more indicative performance metric than others, however this did not translate directly to improvements in classification performance. We conjecture this may be due to the score-invariant nature of the ROC-AUC statistic -- it only captures the probability that a randomly selected real example will rank higher than a randomly selected bogus example, which is independent of the specific real-bogus threshold chosen. We instead opt to use the accuracy score, as this directly maps to the quantity we want to maximise in our model.

Data-based hyperparameters (training set composition, stamp size, data augmentation) are optimised iteratively by hand due to computational constraints. An approximate real-bogus ratio between 1:2 to 1:3 was found to be optimal, with greater values giving better bogus performance at the cost of recovery of real detections -- we opt for 1:2 in the final dataset. The overall dataset size was found to be the biggest determinant of classification accuracy, with larger datasets showing improved performance -- although this increase was subject to diminishing returns with larger and larger datasets. We chose a training set of O($4 \times 10^5$) examples, as this was roughly the largest dataset we could fit into RAM on training nodes -- naturally this could be increased further by reading data from disk on demand, but given CPUs were used for training there was a need to minimise input pipeline latencies as much as possible to compensate. Model performance was found to be relatively insensitive to the ratio of synthetic transients to minor planets, as long as there were at least 10,000 of both in the training set. Using a dataset where 100\% of the real content came from minor planets led to a $\sim 5\%$ drop in the recovery rate of transients on the test set (see Fig. \ref{fig:TPRmagbin}), whereas a 100\% synthetic transient dataset led to a detrimental 15\% decrease in the recovery rate of minor planets, and a 5\% drop on the transient test set. This surprising result implies that combining both minor planets and synthetic transients has a synergistic effect, with the combination providing better performance overall. The specific composition of the final dataset is listed in Table \ref{tab:datadist}, we found a roughly 2:1 minor planet:synthetic transient ratio to provide the correct balance between overall test set performance and sensitivity to astrophysical transients.

\begin{figure}
    \centering
    \includegraphics[width=\linewidth]{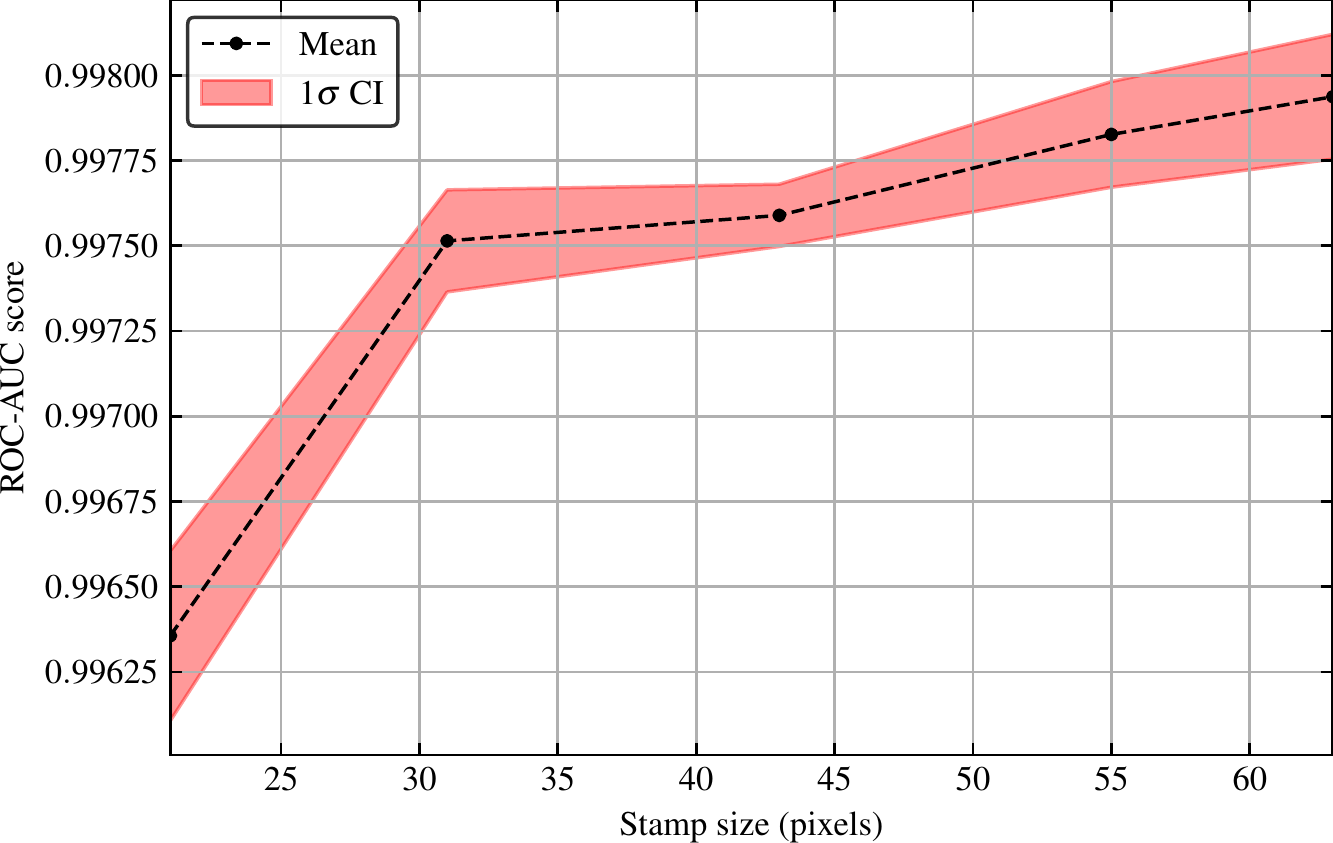}
    \caption{Classifier performance on the test set of a 330,000 example training set as a function of input stamp size. Each point is the average of 3 independent training runs on the same input training set, with the shaded region representing the 1$\sigma$ confidence interval.}
    \label{fig:stampsize_eval}
\end{figure}

A key parameter explored as part of this study is the input stamp size. Larger stamps take longer to generate and more time to perform inference on, so identifying the minimum stamp size possible without affecting performance is crucial. In Figure \ref{fig:stampsize_eval} we show the results of training identical models on an identical 330k-example dataset, with varying stamp size between 21 and 63 pixels. We find that there is no significant increase in performance for our training dataset beyond a stamp size of 55 pixels. The upper limit of this search was set by available RAM, and took 118 hours of compute time to complete.
When scaled through by the ratio of the GOTO/ZTF plate scales (1.4x), our best value of 55 pixels appears remarkably consistent with the 63 pixel stamps that \cite{duev19} found optimal for their network. This is an interesting result, and could imply that the angular scale is actually the more relevant parameter -- this might represent some characteristic length scale that encodes the optimal amount of information about the candidate and surrounding context without including too much irrelevant data.

Network hyperparameters are optimised using the \texttt{Hyperband} algorithm \citep{li17} as implemented in the {\sc Keras-Tuner} package \citep{kerastuner}. This algorithm implements a random search, with intelligent allocation of computational resources by partially training brackets of candidate models and only selecting the best fraction of each bracket to continue training. In testing, this consistently outperformed both naive random search and Bayesian optimisation in terms of final performance.
Table \ref{tab:hyperparam_space} illustrates the region of (hyper)parameter space we choose to conduct our search over. The upper limits for the neuron/filter parameters are set by purely computational constraints -- networks above this threshold take too long to evaluate and train, and so are excluded. We also set an upper limit of 500,000 on the number of model parameters to avoid overly complex models and promote small but efficient architectures. Based on initial experimentation, we require the number of convolutional filters in the second block must be greater than or equal to the number in the first block. This ensures that the largest (and most computationally expensive) convolution operations are performed on tensors that have been max-pooled and thus are smaller, reducing execution time. To maximise performance across all possible deployment architectures, the number of convolutional filters and fully-connected layer neurons are constrained to be a multiple of 8. This is one of the requirements for fully leveraging optimised GPU libraries (such as cuDNN, \citealt{chetlur14_cudnn}), and also enables use of specialised hardware accelerators such as tensor cores in the future. Conveniently, this discretisation also makes the hyperparameter space more tractable to explore.
\begin{table}
    \centering
    \begin{tabular}{l l l l | l}
    {\it continuous} & & & \\
    Hyperparameter & Min & Max & Prior & Selected \\
    \hline
    \hline
    Block 1 filters ($N_1$) & 8 & 32 & linear & 24 \\
    Block 2 filters ($N_2$) & $N_1$ & 64 & linear & 56 \\
    $N_{\text{fc}}$ & 64 & 512 & linear & 208 \\
    Dropout rate & $10^{-2}$ & 0.5 & log & $5.2\times10^{-2}$ \\
    Learning rate & $10^{-5}$ & $10^{-2}$ & log & $6\times10^{-5}$ \\ 
    Regulariser penalty & $10^{-8}$ & $10^{-2}$ & log & $2.0\times10^{-8}$ \\ \\
    
    {\it discrete} & & & \\
    Hyperparameter & \multicolumn{3}{|l|}{Choice} & Selected \\
    \hline
    \hline
    Kernel initialiser & \multicolumn{3}{|l|}{He, Glorot} & Glorot \\
    Kernel regulariser & \multicolumn{3}{|l|}{L1, L2} & L2 \\
    Activation function & \multicolumn{3}{|l|}{ReLU, LeakyReLU, ELU} & LeakyReLU \\
    \end{tabular}
    \caption{Hyperparameter space over which the optimisation search was conducted, split by numerical and categorical variables. The final adopted values are given in the rightmost column.}
    \label{tab:hyperparam_space}
\end{table}

This search took around 1 month to complete on a single 32-core compute node, and sampled 828 unique parameter configurations. The three top-scoring models were then retrained from random initialisation through to early stopping to validate their performance, and confirm that the hyperparameter combination led to stable and consistent results. The top three scoring models achieved accuracies on the hyperparameter validation set of 98.88, 98.64 and 98.54\% respectively. Some of the candidate models had to be pruned from the list due to excessive overfitting.
The best model was then selected based on the minimum test set error. Our final model achieved a test set class-balanced accuracy of $98.72\pm0.02$\% (F1 score $0.9826\pm0.0003$), with the selected hyperparameters listed in Table \ref{tab:hyperparam_space}. This outperforms the version human-tuned by the authors through iterative improvement by 0.6\%, and trains to convergence in around half the number of epochs. We adopt this model architecture going forward, and characterise the overall performance in greater detail in Section \ref{sec:performance}. For this final model, the theoretical maximum ROC-AUC is obtained when the real-bogus threshold is set to 0.4, although in live deployment we opt for a conservative value of 0.8 to minimise contamination.

\subsection{Quantifying classification uncertainty}
Uncertainty estimation in neural networks is an open problem, but is of critical importance for a range of applications. Traditional deterministic neural networks output a single score per class between 0 and 1. This single value would be sufficient to provide a measure of confidence, if properly calibrated. However, neural networks are often regarded as providing over-confident predictions in general, and, worse, providing misidentifications at high confidence. Giving neural networks the ability to make nuanced predictions and account for their own uncertainty in decision making is a potentially powerful improvement, that we discuss in more detail over the next sections.

It is important to be specific and distinguish between epistemic (systematic) and aleatoric (random) uncertainty for the purposes of our classification problem \citep{kendall17_uncertainty}. Aleatoric uncertainty is captured by the classifier's score value, and originates from noise in the input data. More crucial for our application is quantifying the epistemic uncertainty -- that is the uncertainty in our choice of neural network's model weights. This epistemic source of error is directly quantifiable through Bayesian neural networks, and in later sections this is the error, confidence, or uncertainty we refer to and attempt to quantify. In the Bayesian framework, this can be achieved by casting model parameters as probability distributions, and using the mechanics of Bayesian statistics to marginalise the neural network output over these distributions, in the process finding the score posterior. In this way, the uncertainty inherent in model selection can be quantified.  There are various approximate and exact approaches to achieve this which we outline below. 

Dropout \citep{srivastava14} provides a useful form of regularisation in neural networks. At each training step, a fraction $p$ (a tunable hyperparameter) of neuron weights are randomly set to zero, decreasing the effective number of parameters of the model. In this way, overfitting can be prevented and generalisation accuracy can be increased. In traditional neural networks, dropout is not active at inference time so that all neurons are used for making predictions. However, \cite{gal15a_general} demonstrate the profound result that training and evaluating neural networks with dropout is equivalent to performing the approximate Bayesian inference discussed above, with multiple evaluations being equivalent to Monte Carlo integration of the posterior distribution. This is directly applicable to convolutional neural networks, via the Monte Carlo dropout technique (\citealt{gal15b_cnn}; referred to as MCDropout for brevity from now on). 

Alternative approaches to uncertainty estimation exist (Bayes by Backprop, \citealt{blundell15}), which instead directly performs the approximate Bayesian inference by instead casting neuron weights as distributions with associated hyperparameters, then updating these according to the backpropagated gradients (like deterministic NNs). In this work, we opt to use MCDropout for computational efficiency and for maximal compatibility with existing network architectures and software. No changes to the training loop are required, and only a simple wrapper is required at inference to perform multiple predictions with dropout enabled. The only significant additional computational cost for a Bayesian neural network using the MCDropout technique over a deterministic CNN is at inference time, as multiple samples need to be drawn to approximate the posterior. This performance overhead can be mitigated with suitable batching of the dataset.  The ability of neural networks to learn complex, non-linear representations in high-dimensional vector spaces is well-known and utilised throughout machine learning. However, estimation of the uncertainty of products of neural networks is often a barrier to their implementation in scientific applications, where well-grounded determination of errors is important. MCDropout provides a principled way to introduce this.

Although a comparatively new technique, Bayesian neural networks show emerging promise across a variety of astronomical classification and regression tasks -- including supernova light curve classification \citep{moller20}, efficient learning of galaxy morphology \citep{walmsley20}, and age estimation of stars for galactic archaeology \citep{ciucua20}.

There is disagreement in the literature on the precise nature of a Bayesian neural network and how to implement it `properly', from approximate variational inference as used here, to applying some variant of the Markov Chain Monte Carlo sampler over the weight and bias parameters of the neural network. However, what is relevant for the implementation in this work is that examples the classifier is unconfident about are assigned lower confidence scores than obviously real/bogus detections. More complex tests, such as confirming that the classifier's confidence matches the actual confidence of the dataset/some human-derived uncertainty score are beyond the scope of the introductory work presented here.

Whilst these posterior predictions are informative to human vetters, converting them to a single informative summary parameter that captures the overall uncertainty is more useful for integration into pipelines and enabling coarse filtering of candidates. To convert the posterior distributions to meaningful information about the confidence of a given prediction, we utilise the information entropy $\mathbb{H}$. For a binary classification problem, the generic entropy formula can be reduced to:
$$ \mathbb{H} (p) = -p \log_2{p} - (1-p) \log_2{1 - p}$$
where $p$ is the probability of a given detection being real (the real-bogus score).
The entropy is maximised for $p=0.5$, where the probability of being real vs. bogus is equal, or the classifier prediction carries no useful information. We define the classifier confidence $\mathbb{C}$ in terms of the average entropy of the posterior distribution samples, scaling to confidences in the range $[0, 1]$ with the relation
$$ \mathbb{C} = 1 - \frac{1}{N} \sum_{i=1}^{N} \mathbb{H}_i $$
where $N$ is the number of posterior samples and $\mathbb{H}_i$ is the binary entropy of the $i^{\mathrm{th}}$ posterior sample. This metric is equivalent the second term of the BALD acquisition function of \cite{houlsby11}, and is chosen as it is pre-normalised to $[0, 1]$ unlike standard deviation or similar metrics. Naturally the uncertainties we derive here are correlated with the actual output score, but the multiple samples provide sufficient dispersion that this metric is useful to assess model confidence. In future implementations, these raw posterior samples (or some approximating distribution parameters to reduce data needs) could be fed directly into downstream, more specialised classification tools to enable them to make use of the real-bogus classifier's probabilistic predictions in their own score/posterior.

\subsection{Using the uncertainty in classifier predictions} \label{sec:bayesian_tricks}
\begin{figure}
    \centering
    \includegraphics[width=\columnwidth]{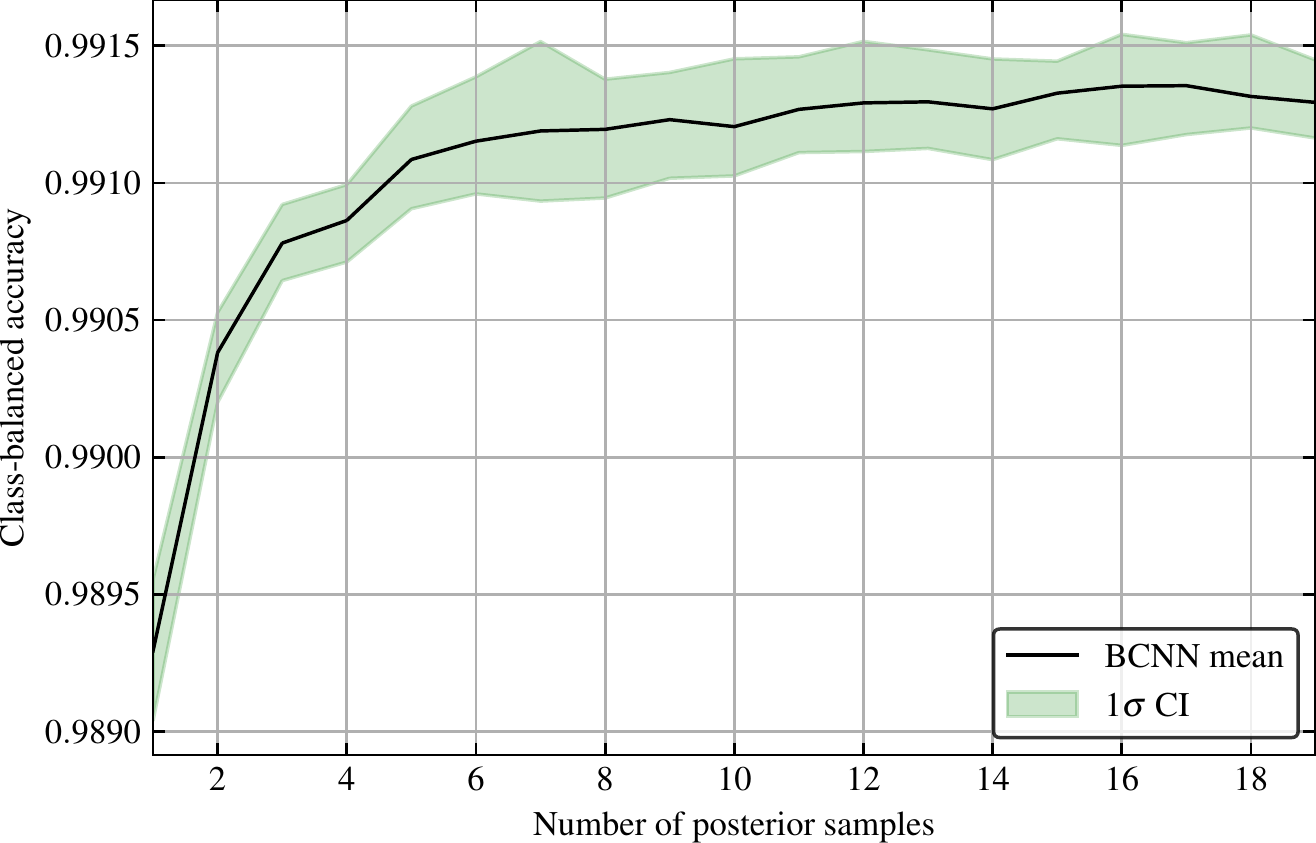}
    \caption{Classification accuracy on the test set from Section \ref{sec:trainingset} as a function of the number of posterior samples averaged. Each point is the average of 10 model runs, with the shaded area corresponding to the $2\sigma$ confidence interval. The BCNN quickly recovers the performance of a deterministic CNN within statistical uncertainty ($99.18\pm0.03$\% accuracy, F1: 0.9877) and provides additional information in the form of confidence.
    No significant improvement in classification accuracy is obtained beyond 10 samples, remaining consistent out to 50 samples.} 
    \label{fig:n_evals}
\end{figure}

One immediate advantage of Bayesian neural networks over deterministic neural networks is the ability to improve classification performance through model ensembling. Figure \ref{fig:n_evals} illustrates the gain in accuracy observed by averaging the predictions of our BNN, as a function of the number of posterior samples. Although small, this is a definite improvement over single-evaluation predictions, and is likely constrained by our dataset. For the majority of positive and negative examples the model is highly confident about the assigned RB score, so averaging over the posteriors does not improve them significantly. This increase in performance is likely to be greater on more complex (multi-class) classification problems, or scenarios where significantly less training data is available.

\begin{figure}
    \centering
    \includegraphics[width=\columnwidth]{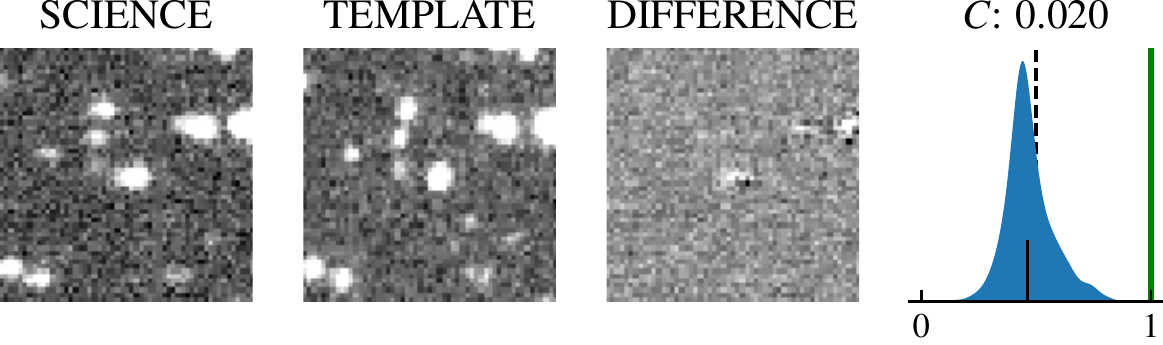}
    \includegraphics[width=\columnwidth]{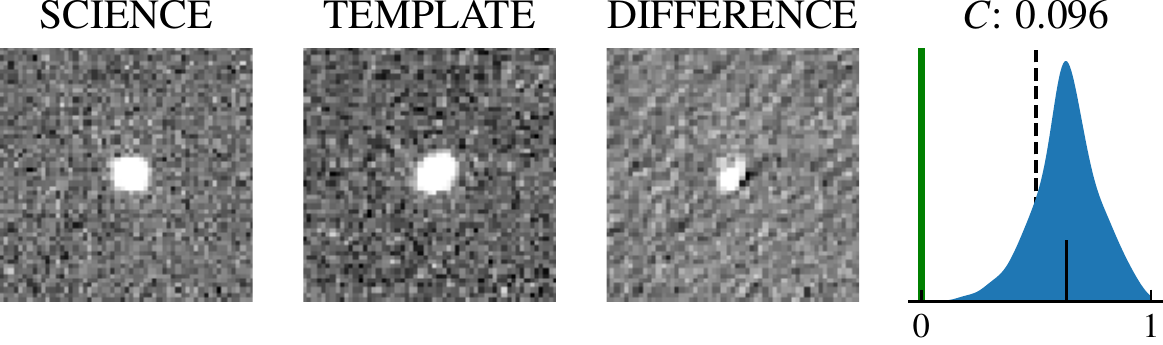}
    \includegraphics[width=\columnwidth]{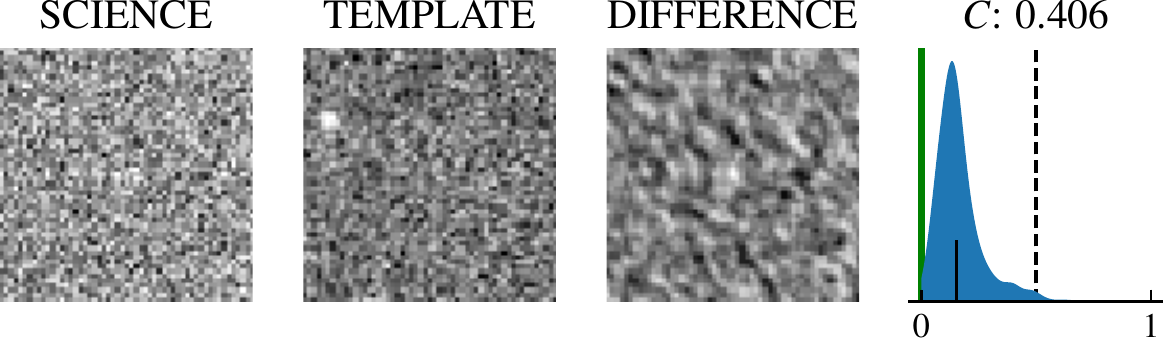}
    \includegraphics[width=\columnwidth]{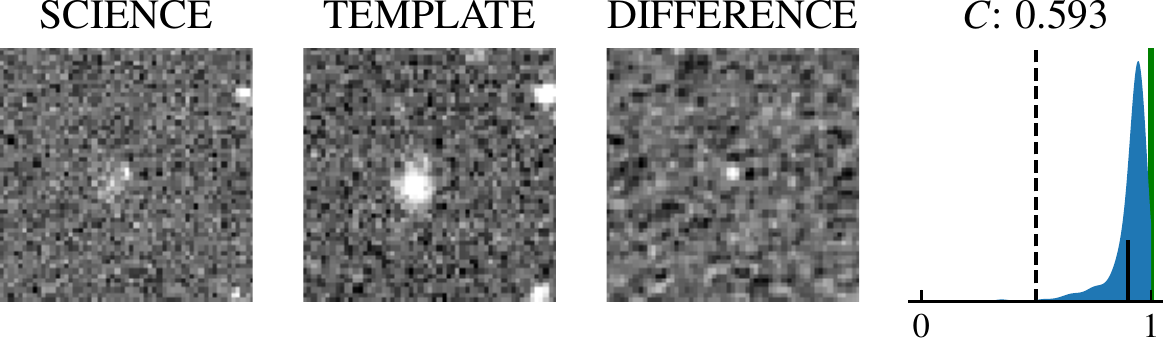}
    \includegraphics[width=\columnwidth]{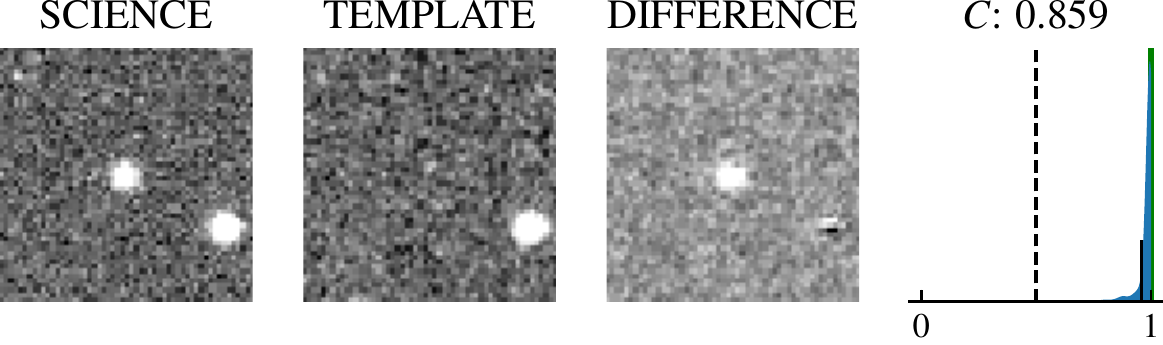}
    \caption{A selection of example posteriors, taken from real GOTO data. The majority of predictions are highly confident, so we select examples of increasing confidence score ($\mathbb{C}$) to display here. Plotted here is a Gaussian kernel-density estimate constructed from 500 posterior samples. The green line indicates the correct label for each candidate, with the black line indicating the mean of the distribution. The dashed line indicates $P_{\mathrm{real}} = 0.5$}.
    \label{fig:exampleposteriors}
\end{figure}

Posteriors and/or associated confidence scores can be added to any downstream candidate evaluation tools, providing an additional metric to inform decisions. Objects with both high score and high confidence are highly likely to be genuine, so can be prioritised in human vetting of candidates. This means more time can be spent looking at more marginal candidates, and obvious detections can quickly be identified. Confidence provides a complementary metric to the pure real-bogus score that can help alleviate some of the issues with the poor dynamic range observed in the classifier outputs at low/high scores. 
Classification is still performed on the consensus real-bogus score derived from the posterior, with the confidence score intended to aid human decision making.
In Figure \ref{fig:exampleposteriors}, we illustrate some example candidates, their associated real-bogus score, and the score posterior.

Classifier confidence is also a useful tool for the training and development process, providing deeper insight into the functioning of the classifier and the associated training set. Predictive uncertainty provides a useful heuristic to clean datasets of mislabelled data. Misclassified detections that the classifier returns a high confidence for are very likely to be mislabelled, as the confidence score is partially based on seeing large numbers of similar detections in the training set. These frames can be actively prioritised in any human relabelling efforts, or fixed cuts on the confidence can be utilised to perform this in a semi-automated way. This `optimal relabelling' scheme provides a method for human vetters and machine learning models to collaboratively and iteratively refine noisy labels. Our label noise is introduced as humans are imperfect judges of real/bogus, and interpret the vetting rubric in different ways leading to inconsistencies which can harm model performance.

We demonstrate the effectiveness of this procedure on the training set built in this work by training the model first on the uncleaned dataset, then attempt to relabel the misclassified detections in the training and test set ordered by decreasing confidence. This amounts to a substantial task of 3580 stamps, which would take a prohibitively long time to relabel by hand, notwithstanding the possibility of human bias in the relabelling. We instead here propose a heuristic re-labelling scheme based on the BALD score of \citet{houlsby11} that leverages the simplistic nature of binary classification. 

The model is first trained on the `unclean' dataset generated with the approaches in Section \ref{sec:trainingset}, then the BALD score is evaluated over the misidentifications in the test and training sets. From here, a new set of labels is derived by flipping the labels of those examples that have a BALD score less than (thus confidence higher than) the median -- effectively accepting the prediction of the classifier over the human vetter. This approach is naturally capable of flipping the labels of accurately labelled stamps incorrectly, but by imposing this cut in classifier confidence it ensures that the majority of relabelled stamps each round correspond to regions of classifier parameter space that are well-covered by the training set and so are classified at high confidence. This method effectively trades active human labelling time for passive background computational time, and can be applied iteratively as suggested above to progressively improve the quality of the dataset labelling.  We manually checked a subset of the sources selected to be re-labelled to verify these were sensible and indeed found they were mislabelled detections that had leaked through the quality cuts we applied. After 1 round of the heuristic relabelling routine outlined above, the class-balanced accuracy achieved on the classifier test set improved markedly from $98.72\pm0.02$ to $99.12 \pm 0.01$\% (F1 score: $0.9826\pm0.0003$ to $0.9877\pm0.0002$), demonstrating the efficacy of this approach. We adopt this cleaned dataset for the following sections.

\begin{figure*}
    \centering
    \includegraphics[width=\linewidth]{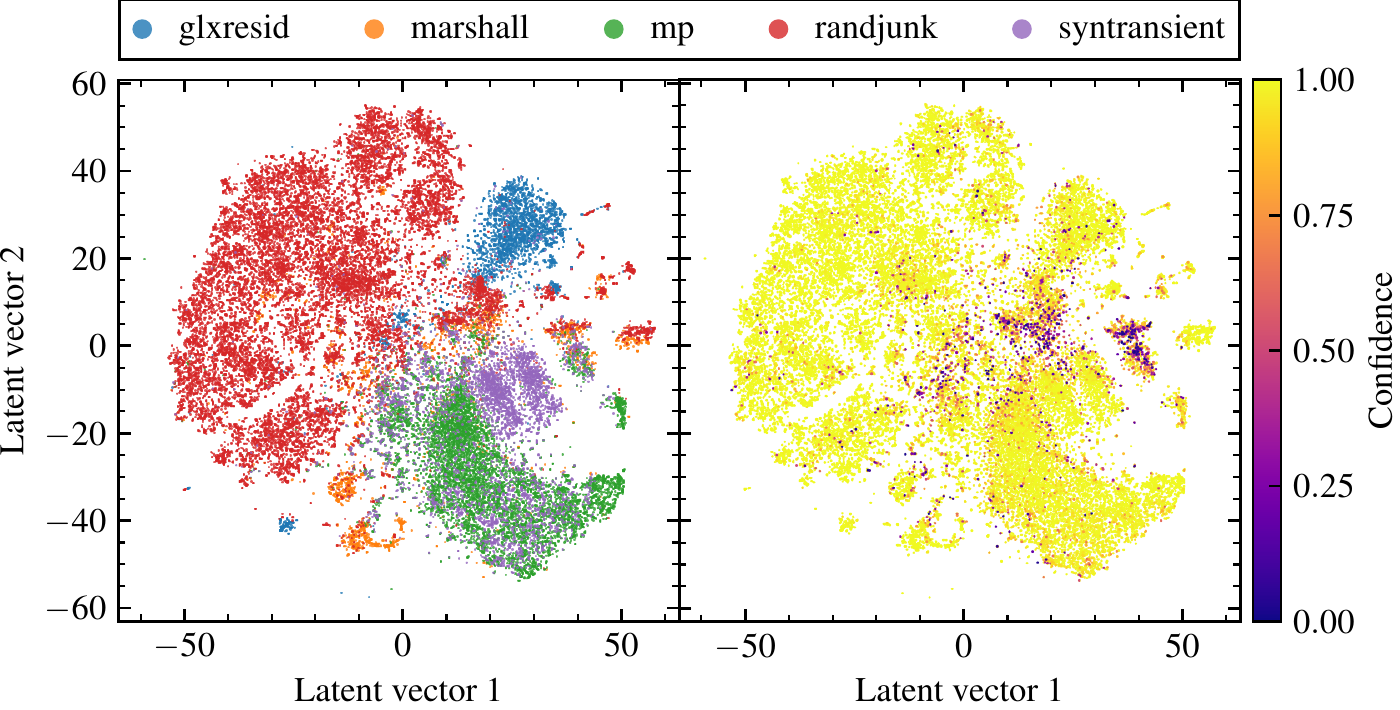}
    \caption{Example class-clustering (left) and confidence (right) maps generated from the classifier's test set. Each colour in the left panel represents a specific sub-class of detections, where colour on the right represents classifier confidence. The top legend gives the classes corresponding to each colour in the left panel. Regions of low confidence in the right panel tend to correspond to cluster boundaries in the left, where there is more uncertainty about which class each example belongs to.}
    \label{fig:tsne_embedding}
\end{figure*}

When visualised in an intuitive way, this confidence score can provide insights into the specific families of detection that the classifier is uncertain about. A natural approach to combine this with is to examine the latent space of the neural network. The first convolutional stage of our network can be thought of as a feature extractor, with the resultant feature vector encoding high-level information about the morphological characteristics of our dataset, providing insight about potential groupings of detection types through clusterings in this space. To explore the latent space within our model, we apply t-stochastic neighbour embedding (t-SNE, \citealt{vandermaaten08}) to the output vector of the layer prior to the fully-connected classification layer to reduce the dimensionality and identify clusterings of common data points. 
The combined process projects an 800-dimensional vector space down to (in our case) a 2D plane. In this space, points with similar latent vectors appear close to each other, thus providing a clustering of the latent space which can be used to visualise the internals of the neural network.
This is a purely diagnostic clustering for visualisation purposes, as t-SNE does not preserve global distances, nor does it provide a bidirectional mapping from the compressed latent space to the full latent vector space. Figure \ref{fig:tsne_embedding} illustrates this technique applied to the test set, coloured by both detection sub-class (left) and classifier confidence (right). 

A useful insight this compressed space provides is the ability to identify clusters of low-confidence points. This immediately reveals types of detection where the classifier may be uncertain, due to intrinsic difficulty of classification (sources close to the detection limit, nuclear transients, unusual PSFs), or scarcity of training data in general. The fact that there are clear divisions between the coloured sub-classes in the left panel of Figure \ref{fig:tsne_embedding} implies that the classifier has learned something about the intrinsic morphology of the detections beyond simple real-bogus division. Neither the classifier nor the clusterer receive these higher-level metalabels, so the clear partitions between the subclasses is purely a result of the internal representations learned.

For more complex datasets where the labelling budget for training examples is limited, Bayesian neural networks enable active learning -- a process where the model identifies input data from a large unlabelled pool that would provide the greatest gain in information to it, using the uncertainty. This has been applied to convolutional neural networks with great success \citep{gal17_activelearn}, and is likely a useful tool for fine-tuning existing training sets in light of new data. We trialled Bayesian active learning as a tool to build the training set presented in this work using the BALD score \citep{houlsby11} as our acquisition function, although it showed no significant improvement over a random selection from the unlabelled pool. This is likely due to the formulation of our classification problem -- using only binary labels, and our data being dominated by large numbers of high-confidence real and bogus examples -- only rare examples which add little to the overall classification accuracy are acquired. The additional complexity introduced by a multi-class labelling scheme along with the greater entropy provided by having multiple output neurons would likely yield better results.

\section{Evaluation of classifier performance} \label{sec:performance}
Machine learning algorithms acquire inherent and often subtle biases based on the training set used in their construction. Given the automated nature of our data set generation, it is particularly important to verify that performance is consistent across a range of parameters of interest, such as transient magnitude. Some care is required in choosing the test set for evaluating classifier performance in a real-world setting, as the training set has been augmented with both human-labelled data and fully synthetic data. Although a low FPR/FNR on the validation and test data is encouraging as it is artificially made more difficult for the classifier to learn, it is not directly representative of the performance we should expect in deployment as a non-negligible component of it is synthetic. Performance characterisation should be reinforced with extensive testing on representative samples of GOTO data. A particular focus is to confirm that the synthetic augmentation scheme we implement leads to genuine improvements in the classifier's recovery rate of real transient detections. We also emphasise that in following sections, we effectively test the performance of the real-bogus classifier in isolation -- the `real-world' detection efficiency is the product of the efficiency of multiple pipeline stages, most crucially image subtraction and source extraction. Exploring the impact of these steps is beyond the scope of this paper, and thus are left to future work.

In the following sections, we use `accuracy' to refer to the class-balanced accuracy, as it is more appropriate for our mildly imbalanced classification task. We also quote results based on the mean scores of 10 posterior samples (motivated by the saturation observed in Figure \ref{fig:n_evals}) since individual evaluations of a Bayesian neural network using MCDropout are based on weaker classifiers due to the presence of dropout. Typical uncertainties (estimated as the standard deviation) on the metrics below are $< 0.05$\%, largely arising from the small number of examples around the decision boundary -- where uncertainties exceed this they are given explicitly.

\subsection{Performance on the test set}
To provide a more granular view of the classifier performance, we further split the test set into two groups for the purposes of evaluation. The first comprises of only the minor planet and random bogus detections. We also test a synthetic transient/galaxy residual test set, to verify that the classifier can genuinely discriminate between galaxies and galaxies with transients. This also reveals any strong performance differences between the two main positive classes, which could skew metrics evaluated on the whole dataset. For both test sets, the human-inspected Marshall data is deliberately excluded, since it is significantly more challenging for the classifier than normal detections and does not accurately reflect the true data distribution.

\begin{figure}
    \centering
    \includegraphics[width=\linewidth]{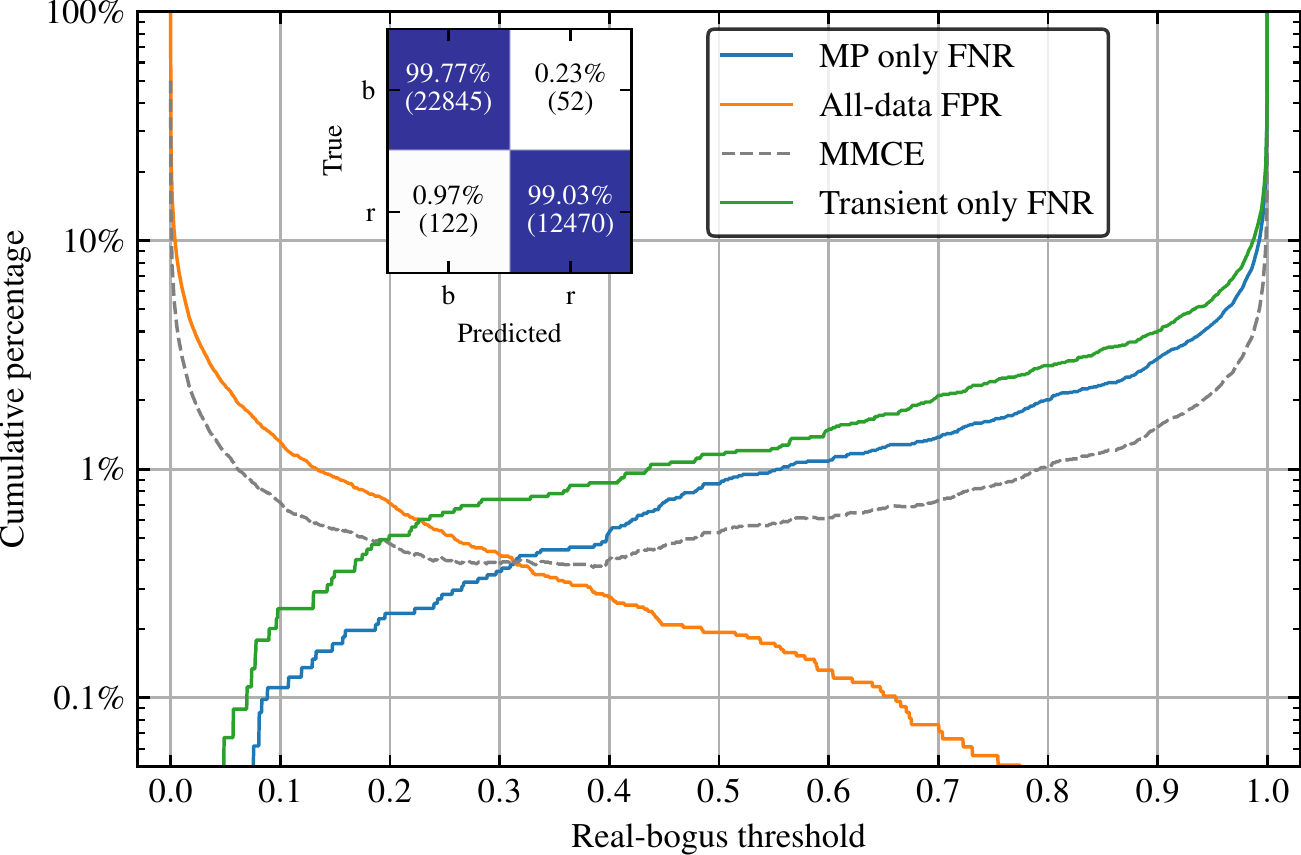}
    \caption{False positive/negative rate evaluated on the test set, excluding Marshall examples. Performance metrics are split based on minor planet and synthetic transients. The grey dashed line (MMCE) represents the full-dataset mean misclassification error, which is below 1\% between real-bogus scores of 0.1 -- 0.6. Inset: confusion matrix, evaluated on the full test set. There is a slight difference in the false negative rates achieved between the minor planets and synthetic transients, reflecting the increased difficulty posed by complex host morphology and subtraction residuals.}
    \label{fig:fprfnr}
\end{figure}

The best-scoring classifier after hyperparameter optimisation shows excellent performance, attaining balanced accuracies of $99.49$\% (F1: 0.9935) and $99.19$\% (F1: 0.9925) on the minor planet and synthetic transient test datasets respectively. Figure \ref{fig:fprfnr} illustrates the false positive and negative rates for the classifier on both the minor planet and transient datasets, as a function of the real-bogus threshold chosen. There is a clear  difference in false negative rate between the minor planet and transient datasets, reflecting the increased difficulty associated with the complex host morphology associated with the transient examples.
The classifier displays a notable skew in the FPR/FNR equality point towards lower values. This is a result of the Marshall injections in the training set, which are made more difficult to learn than the random bogus detections due to being misclassified by the previous classifier. This does not affect classification accuracy, and could be fixed by applying a power transform to the classifier output if required, conditioned on the validation set.

\begin{figure}
    \centering
    \includegraphics[width=\linewidth]{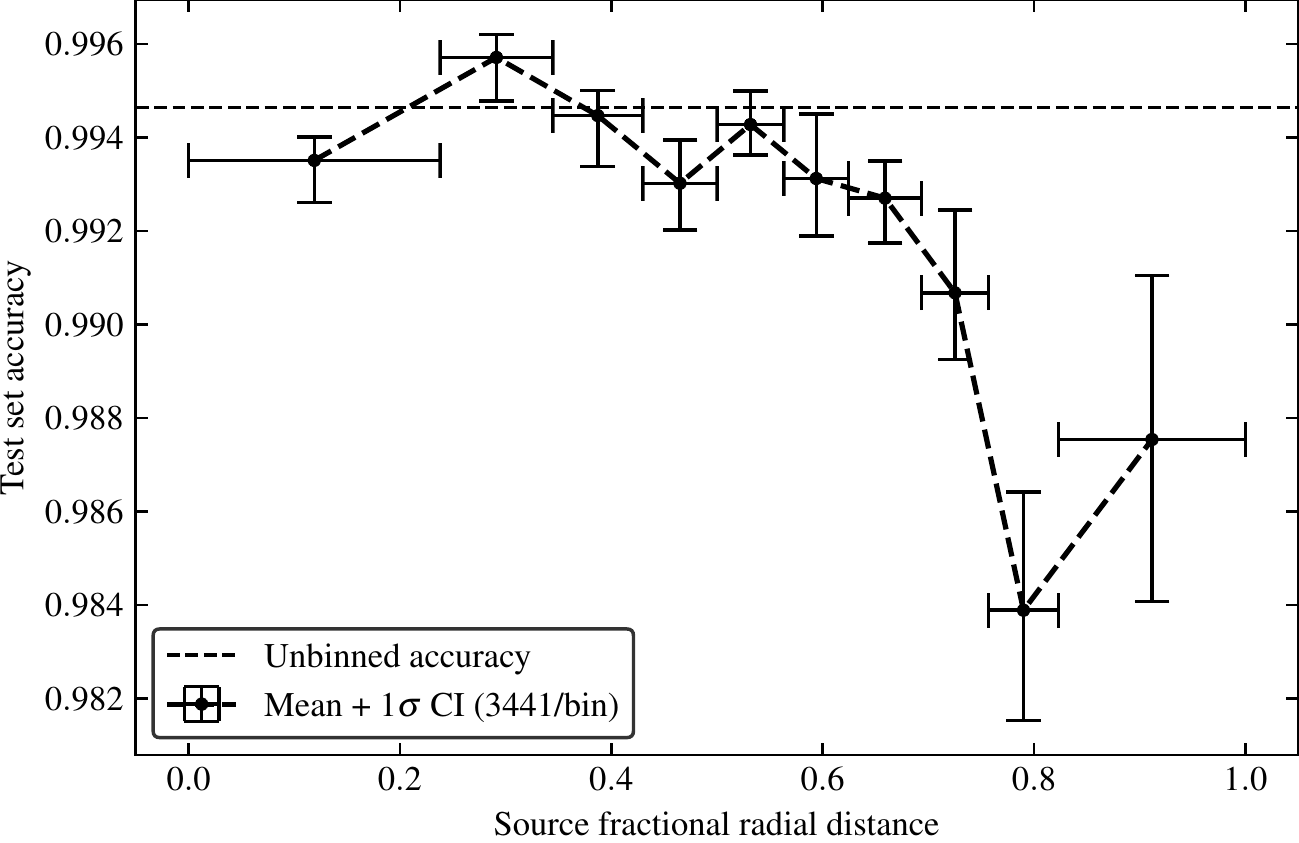}
    \caption{Class-balanced accuracy evaluated on the test set as a function of detector position. We use a series of concentric radial bins, chosen to contain equal numbers of sources for uniform statistics. We scale the radius through by the detector size to give a relative picture of performance. The drop in performance at large radial distances is primarily caused by the extreme optical distortion present in the early GOTO prototype, and only a minor drop of 1\% in accuracy in these challenging conditions demonstrates the very robust performance of our classifier. With the design-specification GOTO optics, we anticipate this curve will be level within error.}
    \label{fig:spatialperf}
\end{figure}

Given the spatially-variable optical characteristics present in the GOTO prototype, it is important to confirm that our classifier provides good performance across the full detector -- and not simply in the centre where distortion is minimal. In Figure \ref{fig:spatialperf} we plot the class-balanced accuracy score as a function of radial position on the detector, using a series of radial bins chosen to equalise source density. These radial bins are scaled through by the maximum value (corresponding to the image corner) to provide a scale-free measurement of detector position. Class-balanced accuracy is used here as the real-bogus fraction varies as a function of detector position, and care must be taken to account for this. We find a consistent performance of $\sim$99\% out to a fractional radial distance of 0.7, with a slight drop of 1\% out at the far edge of the image. This is primarily due to the severe distortion found in the image corners of the GOTO prototype optical tubes, which produces very challenging detections (abnormal PSFs, strong vignetting) both for source extraction and real-bogus classification. Some contribution to this performance decrease is likely from good quality sources close to the edge of the image or close to the edge of the science-template overlap. Estimating reliably these sources and their contribution to the numbers in each bin is a complex task. Suffering only a 1\% decrease in performance in these extremely challenging conditions demonstrates the overall robustness of the classifier. With the significantly improved optical quality of the GOTO design specification OTAs, we anticipate that future versions of our classifier trained on data from the upgraded system will display a constant (within statistical error) classification accuracy as a function of detector position.

\subsection{Performance on spectroscopically confirmed transients} \label{sec:spectrans}
To provide the most accurate assessment of transient-specific classifier performance and further confirm that our algorithmically-generated training set generalises well, we assemble a test set of genuine astrophysical transients. This set was found by cross-matching a list of all spectroscopically confirmed supernovae reported to the Transient Name Server (TNS) since January 2019 with the GOTO master candidate table. Those with an associated GOTO candidate within 3 arcsec, with TNS discovery magnitude greater than the GOTO source magnitude, and only found in GOTO data after the formal TNS discovery date are accepted. With these cuts, purity is favoured over completeness, a deliberate choice to ensure that the test set is as clean of false positives as possible. This yields 877 known transients recovered in the GOTO prototype data. The whole-sample recovery rate is $97.2 \pm 0.3\%$, consistent with the performance achieved on the synthetic transients. This is a strong indicator that our generation algorithm for synthetic transients produces convincing detections which are useful for learning to detect genuine transients. Uncertainties on the TNS-derived set are larger than for our synthetic datasets due to both the smaller sample size and the increased complexity of the real dataset.

\begin{figure}
    \centering
    \includegraphics[width=\linewidth]{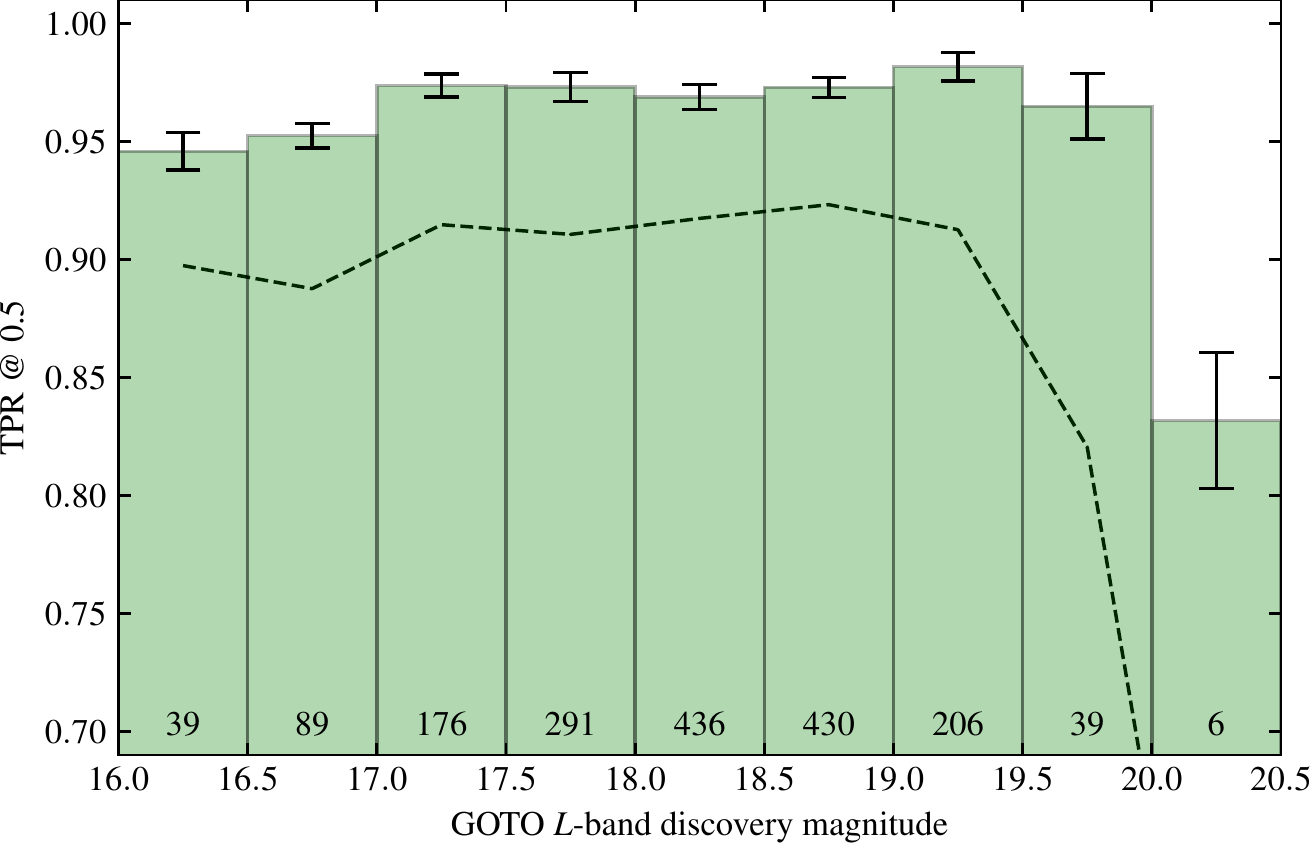} \\
    \vspace{5mm}
    \includegraphics[width=\linewidth]{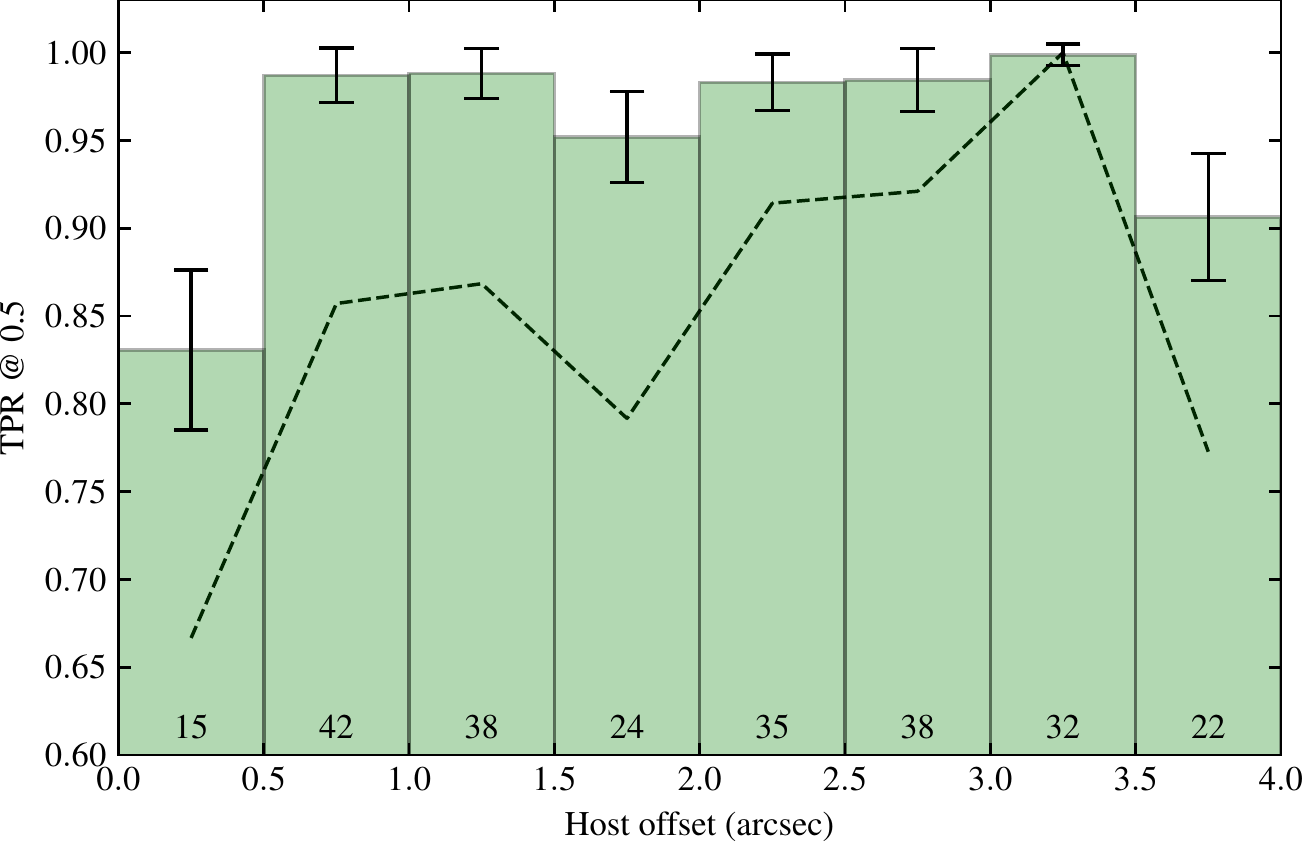}
    \caption{{\it Top panel:} Recovery rate (TPR) as a function of GOTO discovery magnitude, at a fixed real-bogus threshold of 0.5. The dashed line indicates the performance of a classifier with a similarly sized training set, but with only minor planet detections. Error bars are derived directly from the classifier score posteriors. The number of detections per bin is written below each bar. The sharp drop-off in the number of detections beyond $L \sim 19.5$ is associated with the median 5-sigma limiting magnitude of the GOTO prototype, thus expected.
    {\it Bottom panel:} Recovery rate of transients that can be reliably associated with a host galaxy (as cross-matched with WISExSuperCosmos, \citealt{bilicki16_wisexscos}) as a function of host offset. As above, error bars are derived from the classifier score posteriors, and a similarly-sized minor planet-based classifier is plotted for comparison. There is a marked improvement in the recovery rate for very small host offsets, particularly for nuclear transients.
    }
    \label{fig:TPRmagbin}
\end{figure}

To confirm that consistent performance across a wide range of magnitudes is attained, the recovery rate is evaluated across a series of magnitude bins. Figure \ref{fig:TPRmagbin} illustrates the transient recovery rate as a function of GOTO $L$ band magnitude.
We find that the classifier maintains excellent performance across the full magnitude range of detections accessible to GOTO, even towards fainter magnitudes. Our galaxy augmentation scheme provides up to a 30\% improvement in recovery rate at magnitudes fainter than $L \sim $19.5 over a pure minor planet training set. This marked improvement at the faint end of our detection range is powerful, as the expected number of other transients increases as a function magnitude, meaning this improvement in recovery rate will yield a corresponding increase in the total number of transients recovered by GOTO. Of particular relevance for GOTO, we expect the majority of kilonovae within the current GW detection volume to also occupy this magnitude range, increasing significantly our recovery rate of these faint transients in particular.

Our augmentation scheme also provides a significant improvement for sensitivity to nuclear transients, considered to be a more difficult transient morphology to detect. Motivated by the typical RMS astrometric noise level of GOTO images, we adopt a fixed threshold of 0.5 arcsec to distinguish between nuclear and offset transients. We find a $13\pm5$ \% increase in the recovery rate of nuclear transients using the transient-optimised classifier compared to a pure minor planet classifier, on a sample of 15 confirmed detections. This is a direct result of the host offset distribution chosen for the augmentation scheme, which permits full freedom to generate close-in nuclear configurations. The main obstacle to improving this further is the inherent quality of the galaxy subtraction residuals, which limits our bright-end performance.

\subsection{Further characterisation}
Although the main transient sources of interest for GOTO will overwhelmingly be fainter than the saturation level ($L\sim15$), there are still important secondary science Galactic targets as well as rare transients occurring in nearby Local Group galaxies \citep[e.g. SN2014J;][]{fossey14_2014j} which have the potential to brighten beyond the well-sampled regions of our training set.
To simulate these bright transients, GOTO detections of the first 100 minor planets are used. These have magnitudes from $L\sim$10--14, and have well-constrained orbits. Using the {\sc skyfield} code \citep{rhodes19_skyfield}, we generate nightly ephemerides for each minor planet, and locate all difference image detections associated with each object. This yields a benchmark set of around 200 bright asteroid detections. Of the 207 detections, 99.5\% are recovered, showing good consistency with the recovery rate on the fainter minor planets in the classifier test set. Of those minor planets with $L \lesssim 10$, 100\% are recovered, although small-number statistics limits the usefulness of this metric. This bright-end testing demonstrates the excellent dynamic range of the classifier, showing high (>90\%) recovery rates from $10^{\rm th}$ -- $20^{\rm th}$ magnitude.

Through the host offset distribution choice we make, we expect to generate a reasonable number of transients at ~zero offset, so this region of parameter space should not be empty in the training set.
To test the performance in this regime we repeated the procedure outlined in Section 2.2, except with the host offsetting routine disabled to generate synthetic detections overlapping the galaxy nucleus only. This generated ~5,100 synthetic nuclear transients, with a magnitude distribution consistent with that in Figure 1. Testing our model against this dataset (with the negative examples being galaxy residuals as in Section \ref{sec:trainingset}, we obtain a 97.5\% accuracy, with a recovery rate (TPR) of $\sim$ 96\%. These scores are lower than the full-dataset scores, reflecting the increased difficulty of classification in this regime. The average prediction confidence on the real component of this set is 0.9390, which is less than the average prediction confidence on the real members of the test set is 0.9626, reinforcing that these detections are more difficult than the `average' real detection.

\begin{figure}
    \centering
    \includegraphics[width=\linewidth]{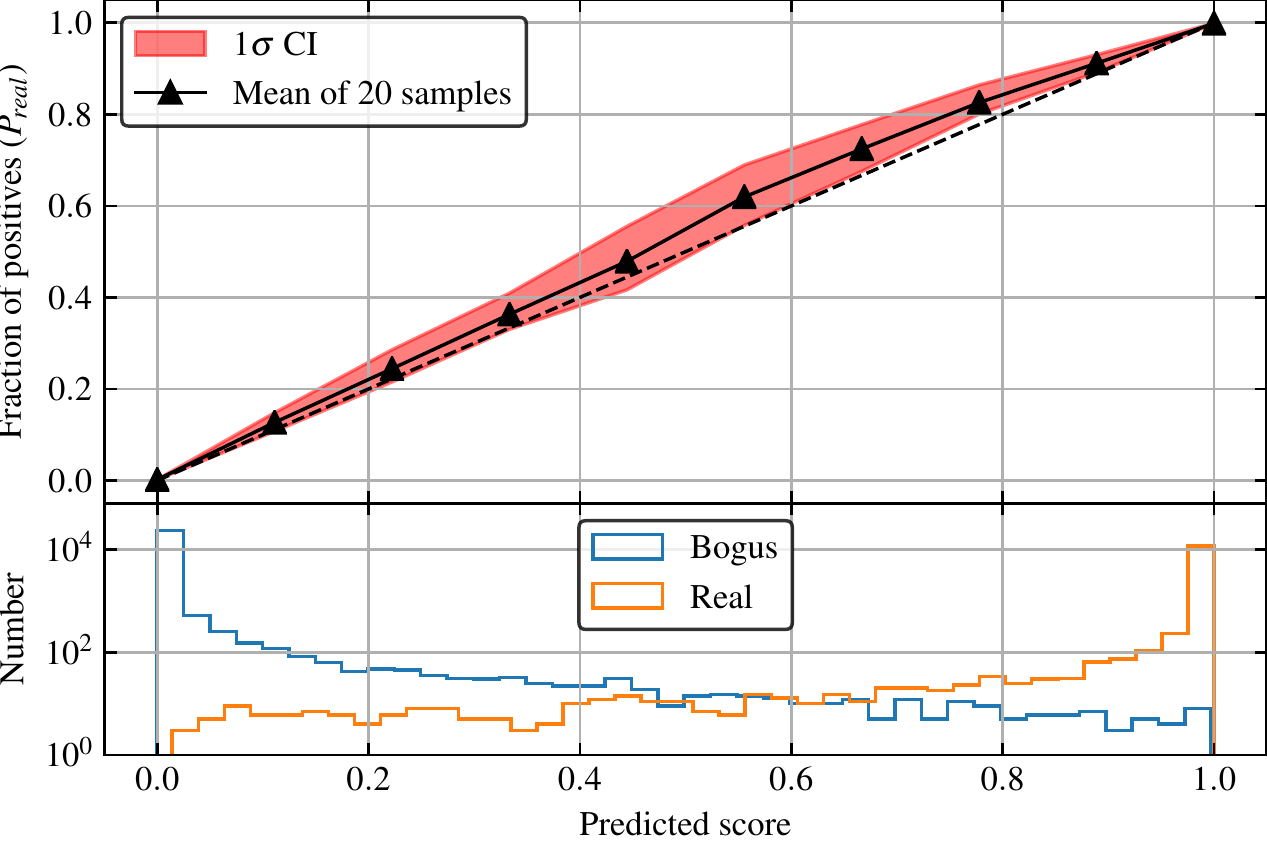}
    \caption{Top panel: classifier calibration curve, illustrating how well the classifier's output score corresponds to probability. The mean of 20 samples and the 1$\sigma$ confidence interval are plotted to show that individual draws from the posterior remain well-calibrated. Bottom panel: Score distribution for both real and bogus examples -- with the relative scarcity of examples with $0.2 < RB < 0.8$ accounting for the greater uncertainty in calibration.}
    \label{fig:calcurve}
\end{figure}

Another important factor to consider with any classifier is how closely the output correlates with probability -- known as calibration. Although this does not necessarily impact on the classification performance, having scores that accurately reflect the probabilities of being real/bogus is important for human use of classification outputs  and is important for performing inference using classifier scores. In Figure \ref{fig:calcurve}, we illustrate the calibration of this classifier by plotting as a function of classifier score the fraction of real detections at a given score. Our uncalibrated classifier shows excellent calibration, and does not show the characteristic sigmoidal calibration curve of other uncalibrated classifiers such as random forests \citep{classifiercalibration15}.
Calibration becomes increasingly important if different machine learning models are chained together, with downstream classifiers using the posterior probabilities of the main real-bogus classifier. With our high degree of calibration, we are justified to use our $RB$ score as a proxy for $P_{real}$ (the probability a given source is real) in such implementations.

One significant benefit of using a Bayesian neural network is a built-in indicator of out-of-distribution data -- that is data poorly represented by or unseen in the training set. For input data that is completely different to the training set, the classifier will return a low confidence score which can then be used to remove/deprioritise the candidate in downstream applications. This confidence can also be used to optimise candidate vetting efforts, with the highest-confidence candidates being a natural choice to prioritise over lower-confidence, lower quality detections.

In principle, the task-specific knowledge encoded in our trained network weights can be used to accelerate the training of similar real-bogus classifiers through transfer learning, and in principle increase generalisation \citep{yosinski14_transferlearning}. This requires that the same data input structure is used and there are no changes to model hyperparameters. However, we caution that training in this way is susceptible to local minima and does not offer the opportunity to change the model hyperparameters that training from scratch does -- in Section \ref{sec:hyperparams} we have demonstrated the sizeable performance improvements doing a full hyperparameter search can yield, and so encourage this.

The techniques and framework we implement in this paper are naturally extensible to more challenging astronomical classification tasks such as those outlined at the end of Section \ref{sec:realbogus_overview}. A key focus is more fine-grained classification -- being able to distinguish variable stars, supernovae, nuclear transients and other astrophysical objects of interest in an automated (and crucially, accurate) way. Figure \ref{fig:tsne_embedding} already hints at this being a fruitful approach, as we see evidence of morphological differentiation in both the positive and negative classes through the emergence of smaller sub-clusters. 
Similarly, leveraging the wealth of contextual information available from astrophysical surveys in a principled, informative, and efficient way within the framework of deep learning poses an open challenge, with potentially significant gains possible. We aim to address these challenges, among others, with development of future generations of the classifier we implement here.

\section{Conclusions}
We demonstrate a data-driven approach to generating large, low-contamination training sets, which along with our novel augmentation scheme can be used to train high-performance, transient-optimised real-bogus classifiers. By combining real PSFs from minor planets with galaxies, we generate realistic synthetic transients that provide a measurable improvement in the recovery of genuine astrophysical transients. This technique is computationally lightweight, easily implemented, and directly applicable to a variety of both current and future transient survey streams/datasets.

We also demonstrate the efficacy of Bayesian neural networks for the first time in real-bogus classification, and demonstrate the unique insights that confidence estimation can bring to the real-bogus problem. Being able to assign epistemic confidences to classifier predictions in addition to the more typical real-bogus score provides another parameter for human vetters further downstream to use in identifying promising candidate detections -- this can potentially be used in future to further automate decision making in the context of follow-up and reporting. Techniques such as this that minimise human involvement in data-gathering and labelling will become increasingly important in the new `big-data' era of astronomy that large-scale projects such as the Rubin Observatory and SKA will bring about. 

Our classifier demonstrates excellent performance across a wide magnitude range, with a missed detection rate of 0.5\% at a fixed 1\% false positive rate, and up to 30\% improvement in recovery rate of astrophysical transients in the challenging faint end. This has the potential to markedly increase the number of faint transients GOTO can discover, and significantly improves the prospects for detecting the kilonova afterglows of gravitational-wave driven mergers GOTO was designed to find. We anticipate that improvements to the quality and stability of GOTO's hardware and dataflow will bring significant performance gains for the real-bogus classifier presented here. 

GOTO is due to undergo significant expansion over the coming years, with a final configuration of 4 installations spread across a northern (La Palma) and southern (Siding Spring) site providing a high-cadence datastream covering almost the whole sky down to 20th magnitude every 2-3 days. The tools developed in this work have generated a classifier that is capable of handling and sifting the accompanying volume of candidate transient detections with robust accuracy and high sensitivity.

\section*{Acknowledgements}
We thank the anonymous referee for their insightful comments which helped improve the quality of this manuscript.
The Gravitational-wave Optical Transient Observer (GOTO) project acknowledges the support of the Monash-Warwick Alliance; Warwick University; Monash University; Sheffield University; the University of Leicester; Armagh Observatory \& Planetarium; the National Astronomical Research Institute of Thailand (NARIT); the University of Turku; the University of Manchester; the University of Portsmouth; the Instituto de Astrof\'{i}sica de Canarias (IAC) and the Science and Technology Facilities Council (STFC).
DS, KU, BG and JDL acknowledge support from the STFC via grants ST/T007184/1, ST/T003103/1 and ST/P000495/1.
JDL acknowledges support from a UK Research and Innovation Fellowship (MR/T020784/1).
RPB, MRK and DMS acknowledge support from the ERC under the European Union’s Horizon 2020 research and innovation programme (grant agreement No. 715051; Spiders).
POB and RS acknowledge support from the STFC.

This research made use of Astropy,\footnote{\url{http://www.astropy.org}} a community-developed core Python package for Astronomy \citep{astropy13, astropy18}, and scikit-learn \citep{scikit-learn}. 
The resources to support \texttt{astorb.dat} were originally provided by NASA grant NAG5-4741 (PI E. Bowell) and the Lowell Observatory endowment, and more recently by NASA PDART grant NNX16AG52G (PI N. Moskovitz).
This research has made use of IMCCE's SkyBoT VO tool.
This research has made use of data and/or services provided by the International Astronomical Union's Minor Planet Center. 

\section*{Data Availability}
The {\sc gotorb} code is made freely available at \url{https://github.com/GOTO-OBS/gotorb}, along with validation examples for testing. Accompanying observational data used in this work will be made available via upcoming GOTO public data releases.



\bibliographystyle{mnras}
\bibliography{refs} 







\bsp	
\label{lastpage}
\end{document}

%% file: new_nn_annotated.pdf_tex
\begingroup%
  \makeatletter%
  \providecommand\color[2][]{%
    \errmessage{(Inkscape) Color is used for the text in Inkscape, but the package 'color.sty' is not loaded}%
    \renewcommand\color[2][]{}%
  }%
  \providecommand\transparent[1]{%
    \errmessage{(Inkscape) Transparency is used (non-zero) for the text in Inkscape, but the package 'transparent.sty' is not loaded}%
    \renewcommand\transparent[1]{}%
  }%
  \providecommand\rotatebox[2]{#2}%
  \newcommand*\fsize{\dimexpr\f@size pt\relax}%
  \newcommand*\lineheight[1]{\fontsize{\fsize}{#1\fsize}\selectfont}%
  \ifx\svgwidth\undefined%
    \setlength{\unitlength}{745.35040283bp}%
    \ifx\svgscale\undefined%
      \relax%
    \else%
      \setlength{\unitlength}{\unitlength * \real{\svgscale}}%
    \fi%
  \else%
    \setlength{\unitlength}{\svgwidth}%
  \fi%
  \global\let\svgwidth\undefined%
  \global\let\svgscale\undefined%
  \makeatother%
  \begin{picture}(1,0.3194908)%
    \lineheight{1}%
    \setlength\tabcolsep{0pt}%
    \put(0,0){\includegraphics[width=\unitlength,page=1]{new_nn_annotated.pdf}}%
    \put(0.00247592,0.00325062){\color[rgb]{0,0,0}\makebox(0,0)[lt]{\lineheight{1.25}\smash{\begin{tabular}[t]{l}conv\end{tabular}}}}%
    \put(0.09371274,0.00325062){\color[rgb]{0,0,0}\makebox(0,0)[lt]{\lineheight{1.25}\smash{\begin{tabular}[t]{l}conv\end{tabular}}}}%
    \put(0.20393697,0.00334103){\color[rgb]{0,0,0}\makebox(0,0)[lt]{\lineheight{1.25}\smash{\begin{tabular}[t]{l}pool (2x2)\end{tabular}}}}%
    \put(0.33147888,0.04287572){\color[rgb]{0,0,0}\makebox(0,0)[lt]{\lineheight{1.25}\smash{\begin{tabular}[t]{l}conv\end{tabular}}}}%
    \put(0.44019762,0.04287572){\color[rgb]{0,0,0}\makebox(0,0)[lt]{\lineheight{1.25}\smash{\begin{tabular}[t]{l}conv\end{tabular}}}}%
    \put(0.53895334,0.04296614){\color[rgb]{0,0,0}\makebox(0,0)[lt]{\lineheight{1.25}\smash{\begin{tabular}[t]{l}pool (4x4)\end{tabular}}}}%
    \put(0.68248415,0.04137423){\color[rgb]{0,0,0}\makebox(0,0)[lt]{\lineheight{1.25}\smash{\begin{tabular}[t]{l}flatten\end{tabular}}}}%
    \put(0.83217848,0.03629036){\color[rgb]{0,0,0}\makebox(0,0)[lt]{\lineheight{1.25}\smash{\begin{tabular}[t]{l}dense\end{tabular}}}}%
    \put(-0.00091516,0.30879691){\color[rgb]{0,0,0}\makebox(0,0)[lt]{\lineheight{1.25}\smash{\begin{tabular}[t]{l}(55,55,4)\end{tabular}}}}%
    \put(0.09754559,0.30073454){\color[rgb]{0,0,0}\makebox(0,0)[lt]{\lineheight{1.25}\smash{\begin{tabular}[t]{l}(53,53,24)\end{tabular}}}}%
    \put(0.22665723,0.29309695){\color[rgb]{0,0,0}\makebox(0,0)[lt]{\lineheight{1.25}\smash{\begin{tabular}[t]{l}(51,51,24)\end{tabular}}}}%
    \put(0.32509758,0.21926665){\color[rgb]{0,0,0}\makebox(0,0)[lt]{\lineheight{1.25}\smash{\begin{tabular}[t]{l}(25,25,56)\end{tabular}}}}%
    \put(0.4379646,0.20568867){\color[rgb]{0,0,0}\makebox(0,0)[lt]{\lineheight{1.25}\smash{\begin{tabular}[t]{l}(23,23,56)\end{tabular}}}}%
    \put(0.55168019,0.19041346){\color[rgb]{0,0,0}\makebox(0,0)[lt]{\lineheight{1.25}\smash{\begin{tabular}[t]{l}(21,21,56)\end{tabular}}}}%
    \put(0.77317104,0.16156024){\color[rgb]{0,0,0}\makebox(0,0)[lt]{\lineheight{1.25}\smash{\begin{tabular}[t]{l}(1400,)\end{tabular}}}}%
    \put(0.86677006,0.14458774){\color[rgb]{0,0,0}\makebox(0,0)[lt]{\lineheight{1.25}\smash{\begin{tabular}[t]{l}(208,)\end{tabular}}}}%
    \put(0.96848413,0.1208263){\color[rgb]{0,0,0}\makebox(0,0)[lt]{\lineheight{1.25}\smash{\begin{tabular}[t]{l}(1,)\end{tabular}}}}%
  \end{picture}%
\endgroup%